\documentclass[12pt, a4paper]{article}
\usepackage[text={17.55cm,24.5cm}, top=3cm, left=2cm]{geometry}  
\usepackage{amsmath,amsfonts,latexsym,mathtools}
\usepackage{mathrsfs} 
\usepackage{graphicx, subfigure}
\usepackage[hidelinks]{hyperref} 
\numberwithin{equation}{section}

\usepackage{xcolor}

\usepackage[title, titletoc, page]{appendix}

\usepackage[super]{nth}

\usepackage{comment}

\usepackage{tabularx}
\usepackage{hhline}
\newcolumntype{Y}{>{\centering\arraybackslash}X}
\usepackage{array, multirow}

\allowdisplaybreaks[1]

\newcommand{\lb}{\left (}
\newcommand{\rb}{\right )}

\newcommand{\lsq}{\left [}
\newcommand{\rsq}{\right ]}

\newcommand{\dd}[1]{\; \mathrm{d} #1}

\begin{document}

\baselineskip=15pt
\vspace{-2.5cm}
\title{Boussinesq-Klein-Gordon and Ostrovsky equations: \\ evolution of cnoidal waves with local defects}
\date{}
\maketitle
\vspace{-22mm}
\begin{center}
{\bf Korsarun Nirunwiroj, Dmitri Tseluiko, Karima Khusnutdinova}
\footnote{Corresponding author: K.Khusnutdinova@lboro.ac.uk} \\[2ex]
Department of Mathematical Sciences, Loughborough University, \\
Loughborough LE11 3TU, UK.\\[3ex]
{\it Dedicated to Lev Ostrovsky on the occasion of his 90th Birthday.}
\vspace{4mm}

\end{center}

\abstract{The Boussinesq--Klein--Gordon (BKG) equation has emerged in the studies of nonlinear bulk strain waves in layered solid waveguides. The bi-directional weakly-nonlinear solution leads to two copies of the Ostrovsky equation, for the right- and left-propagating waves. Importantly, the derivation avoids the so-called `zero-mean contradiction'  between the type of initial conditions in the parent equation and in the reduced model. In this paper, we apply the solution to describe the evolution of cnoidal waves with local periodicity defects and generic localised perturbations, and compare the results with the direct numerical simulations for the full BKG equation. The cnoidal waves with the periodicity defects discussed in our work constitute generalised travelling waves of the Korteweg--de Vries equation, while the Ostrovsky equation leads to a strong burst  (and may lead to a rogue wave), qualitatively similar to the wavepacket emerging from a soliton initial condition, but appearing much faster. We compare the weakly-nonlinear solution with the direct numerical simulations within the bi-directional setting of the BKG equation and show that the  discussed uni-directional waves and evolution scenarios remain stable in the presence of counter-propagating perturbations. }
\bigskip


\newpage

\maketitle

\section{Introduction}
The Ostrovsky equation
\begin{equation}
\lb f_t + \nu f f_x + \mu f_{xxx} \rb_x = \lambda f,
\label{Oe}
\end{equation}
with constant coefficients and subscripts denoting partial derivatives, belongs to the class of universal models of the modern nonlinear wave theory.  The first appearance of the model was related to the description  of the long weakly-nonlinear surface and internal gravity waves in a rotating ocean \cite{O1978} (see also \cite{G1985}).  In this context,  $f$ denotes the amplitude of a dominant linear long-wave mode in the reference frame moving with its linear long-wave speed, and $\nu$, $\mu$ and $\lambda$ are the nonlinearity, dispersion and rotation coefficients, respectively. Later, the Ostrovsky equation also emerged in the studies of the magneto-acoustic waves in a rotating plasma \cite{OS1998} and bulk strain waves in layered elastic solids  \cite{KSZ2009, KM2011}. Scattering of solitary waves in delaminated bi-layers with soft adhesive bonding has been studied in \cite{KT2017,  KT2022, TCPT2024, TCT2025}, using both direct and semi-analytical (weakly-nonlinear) approaches, leading to the single and coupled Ostrovsky equations. The coupled Ostrovsky equations for a pair of strongly-interacting (propagating with close speeds) waves were first derived  in \cite{KM2011}, in the context of waves in solids.  Then, coupled Ostrovsky equations for the gravity waves in stratified fluids on a depth-dependent parallel shear flow supported by suitable body forces were derived in \cite{AGK2013}. The single-mode reduction of the model in \cite{AGK2013} provides the coefficients of the Ostrovsky equation (\ref{Oe}) for surface and internal waves in the presence of an arbitrary depth-dependent parallel shear flow. In the oceanic setting, and in the absence of currents, $\lambda \mu > 0$ (normal dispersion). Examples of shear flows leading to $\lambda \mu < 0$  (anomalous dispersion) were found in \cite{AGK2014}. Extensions accounting for higher-order nonlinearities and weak dependence on the transverse coordinate in the oceanic setting were developed  in  \cite{GPTK2010, NSS2015, GSM2015, ORS2018, G2015}. 

The Ostrovsky equation (\ref{Oe}) implies that for any regular periodic solution on a finite interval $[-L, L]$ (including the case of localised solutions on a large but finite interval)  the mean, or related mass, is zero for any $t\ge 0$:
\begin{equation}
\frac{1}{2L} \int_{-L}^L f \dd{x} = 0.	 
\end{equation}
The original problem formulations do not impose similar constraints on the initial conditions of the related field variables, leading to the so-called `zero-mean contradiction'. This contradiction was avoided in the construction of weakly-nonlinear solutions leading to the Ostrovsky equation in the contexts of the Boussinesq--Klein--Gordon and coupled Boussinesq equations in \cite{KMP2014, KT2019, KT2022} (see also the references therein) by deriving the Ostrovsky equation for the  deviation from the evolving mean fields. Importantly, evolution of the mean fields was described by independent equations. This is not the case in the derivations related to the oceanic setting, where the mean fields and deviations from them remain coupled. Recently, in \cite{NTK2025}, we managed to bypass the zero-mean contradiction in the context of the  rotation-modified Miyata--Maltseva--Choi--Camassa (MMCC-$f$) model developed in \cite{H2007} (see also \cite{Mi1988, Ma1989, CC1996, CC1999}), where the symbol $f$ in MMCC-$f$ stands for the Coriolis parameter.  A large class of uni-directional solutions was constructed by simultaneously expanding the mean fields and deviations from them in powers of $\sqrt{\alpha}$, where $\alpha$ is a small amplitude parameter, and using two slow-time variables, $\tau = \sqrt{\alpha} t$ and $T = \alpha t$, rather than just single slow time $T$, as in classical derivations of the KdV--Ostrovsky--type equations. The resulting Ostrovsky equations satisfy the zero-mean constraint by construction.

There have been intensive studies of the evolution of the waves described by the Ostrovsky equation (\ref{Oe}) with the natural initial conditions corresponding to the solitons of the Korteweg--de Vries (KdV) equation, which is recovered when $\lambda \to 0$  (see \cite{GOSS1998, GH2012, S2020} and references therein). In the oceanic setting, in the absence of currents,  $\lambda \mu > 0$ (normal dispersion), and there are no solitary waves \cite{L1981, GS1991, GHO1998}. The long-time outcome of the evolution is a localised wave packet associated with the extremum of the group velocity  \cite{GH2008, GHJ2013}. The emergence of the wavepacket associated with the extremum of the group velocity was also noted in \cite{YK2001}, in an independent study devoted to the waves in a Toda chain on an elastic substrate, which can be related to the Ostrovsky equation.  Anomalous dispersion  ($\lambda \mu < 0$) can appear in the context of waves in plasma \cite{OS1998} and for oceanic waves propagating over certain shear flows \cite{AGK2014}, leading to the emergence of steady wave packets (solitons) associated with the extremum of the phase velocity.  There were also important studies of the modulational instability of slowly-modulated monochromatic waves  \cite{WJ2014, WJ2017,  JOP2025}. Contrary to the waves described by the KdV equation, there exists a critical wavenumber $k_c$, such that waves with $k>k_c$ are modulationally unstable. Importantly, the long-time outcome of the evolution in modulationally unstable cases is qualitatively similar to that for solitons -- there appears a slower moving wavepacket \cite{WJ2017}, indicating a possible universal role of the modulational instability in the formation of such wavepackets.

Motivated by available observational data for internal waves, in \cite{NTK2025} we studied the effect of rotation on the evolution of KdV cnoidal waves close to the solitonic limit and having periodicity defects and localised perturbations, within the scope of the uni-directional weakly-nonlinear solutions derived from the MMCC-$f$ model. The expansion defect was introduced by cutting a KdV cnoidal wave at the through and inserting a piece of a straight line. We showed that such functions satisfied all (infinitely many) conservation laws of the KdV equation (appropriately understood), the Weierstrass-Erdmann corner condition for broken extremals of the associated variational problem, and a natural weak formulation. Hence,  these functions are generalised travelling waves of the KdV equation, using the terminology introduced in \cite{GNST2020,GS2022,GNS2024}. Being smoothed in pseudospectral simulations, these solutions behaved very closely to travelling waves, with no visible evolution. Moreover, such defects were found to naturally emerge in the evolution of cnoidal waves with localised perturbations, alongside bright and dark breather on a cnoidal wave solutions \cite{KM1975, HMP2023}.
 
 In our present paper, we extend this study to the simplest bi-directional setting of the Boussinesq--Klein--Gordon equation
 \begin{equation}
u_{tt} - c^2 u_{xx} = \varepsilon \lsq \frac{\alpha}{2} \lb u^2 \rb_{xx} + \beta u_{ttxx} - \gamma u \rsq.
\label{BousOst}
\end{equation}
This equation has emerged in studies of long nonlinear longitudinal bulk strain waves in layered elastic waveguides with soft (imperfect) interfaces \cite{KSZ2009, KM2011}, in the limit in which one layer has much greater inertia than the other, and can be treated as a substrate, similarly to the case of a Toda chain on an elastic substrate in \cite{YK2001}. Here, $u$ describes the longitudinal strain, $c$ is the linear longitudinal wave speed, $\alpha$ and $\beta$ are the nonlinearity and dispersion coefficients, respectively, and $\gamma>0$ is determined by elastic properties of the bonding layer or an elastic substrate. We note that up to a scaling of variables one can assume that $c = \alpha = \beta = 1$, but keeping the constants in the model is preferable from the viewpoint of applications.  Boussinesq-type equations generally play an important role in the studies of waves in solids  (see \cite{Mau1999, S2001, P2003, KS2008, EST2011, PTE2017} and references therein).

The paper is organised as follows. We briefly overview the construction of  a weakly-nonlinear d'Alembert-type solution of the Cauchy problem for the BKG equation 
in Section 2. In Section 3, we consider uni-directional solutions with initial conditions corresponding to a cnoidal wave of the KdV equation, and compare the weakly-nonlinear solution with direct numerical simulations for the BKG equation both for $\gamma = 0$ (KdV regime) and $\gamma \ne 0$ (Ostrovsky regime).  In Section 4, we discuss the recently introduced generalised travelling waves of the KdV equation in the form of cnoidal waves with expansion and contraction defects (see \cite{NTK2025}) and perform similar comparisons for the initial conditions with expansion defects. Finally, we model the evolution of the initial conditions in the form of cnoidal waves with generic localised perturbations in Section 5 and make concluding remarks in Section 6.

\section{Bi-directional weakly-nonlinear solution}
\label{sec:2}

Following \cite{KMP2014, KT2019}, we consider the Cauchy problem for the BKG equation on the domain $\Omega = [-L, L] \times [0, T]$:
\begin{align}
&u_{tt} - c^2 u_{xx} = \varepsilon \left[ \frac{\alpha}{2} (u^2)_{xx} + \beta u_{ttxx} - \gamma u \right],
\label{BousOstOld1} \\
&u |_{t=0} = F(x), \quad u_t |_{t=0} = V(x),
\label{BousOstIC1}
\end{align}
where $F$ and $V$ are sufficiently smooth $2L$-periodic functions, and both functions $F(x)$ and $V(x)$ may have non-zero mean values:
\begin{equation}
F_{0} = \frac{1}{2L} \int_{-L}^{L} F(x) \,dx \quad \text{ and } \quad V_{0} = \frac{1}{2L} \int_{-L}^{L} V(x) \,dx.
\label{MeanValIC1}
\end{equation}
Then, from (\ref{BousOstOld1}), (\ref{BousOstIC1}), the evolution of the mean value of $u$ is given by
\begin{equation}
	\langle u \rangle (t) := \frac{1}{2L} \int_{-L}^{L} u(x,t) \dd{x} = F_{0} \cos{\lb \sqrt{\varepsilon \gamma} t \rb} + V_{0} \frac{\sin{\lb \sqrt{\varepsilon \gamma} t \rb}}{\sqrt{\varepsilon \gamma}}.
	\label{MeanVal1}
\end{equation}
The initial-value problem for the deviation from the oscillating mean value $\tilde{u} = u - \langle u \rangle (t)$ has the form
\begin{align}
	\tilde u_{tt} - c^2 \tilde u_{xx} = \varepsilon &\lsq \alpha \lb F_0 \cos{\lb \omega t  \rb}  + \frac{1}{\sqrt{\varepsilon}} \frac{V_0}{\sqrt{\gamma}} \sin {\lb \omega t \rb} \rb \tilde u_{xx} \right. \notag \\
	&\left. ~+ \frac{\alpha}{2} \lb \tilde u^2 \rb_{xx} + \beta \tilde u_{ttxx} - \gamma \tilde u \rsq,
	\label{BousOstEq1}
\end{align}
and
\begin{equation}
\tilde u |_{t=0} = F(x) - F_0, \quad \tilde u_t |_{t=0} = V(x) - V_0,
\label{BousOstICnew1}
\end{equation}
where $\omega = \sqrt{ \varepsilon \gamma}$.

Next, we look for a weakly-nonlinear solution in the form of a bi-directional asymptotic multiple-scale expansion
\begin{align}
	\tilde u \lb x, t \rb &= f^{+} \lb \xi_{+}, \tau, T \rb +  f^{-} \lb \xi_{-}, \tau, T \rb + \sqrt{\varepsilon} P \lb \xi_{-}, \xi_{+}, \tau, T \rb + O \lb \varepsilon \rb,
	\label{WNLSol1}
\end{align}
where 
\begin{equation*}
\xi_{\pm} = x \pm c t, \quad \tau = \sqrt{\varepsilon} t, \quad T = \varepsilon t,
\end{equation*}
and all terms are periodic in $x$ and have zero mean.
(For derivations including higher-order terms see \cite{KT2019}.)

The first non-trivial equation appears at $O \lb \sqrt{\varepsilon} \rb$:
\begin{equation}
- 4 c^2 P_{\xi_- \xi_+} = 2 c f^-_{\xi_- \tau} - 2 c f^+_{\xi_+ \tau} + \frac{\alpha V_0}{\sqrt{\gamma}} \sin \lb \sqrt{\gamma} \tau \rb \lb f^-_{\xi_- \xi_-} + f^+_{\xi_+ \xi_+} \rb.
\label{Peq}
\end{equation}
Averaging with respect to $x$ at constant $\xi_-$ or $\xi_+$ yields the equations
\begin{equation}
2 c f^-_{\xi_- \tau} + \frac{\alpha V_0}{\sqrt{\gamma}} \sin \lb \sqrt{\gamma} \tau \rb f^-_{\xi_- \xi_-} = 0,
\label{f-}
\end{equation}
and
\begin{equation}
2 c f^+_{\xi_+ \tau} - \frac{\alpha V_0}{\sqrt{\gamma}} \sin \lb \sqrt{\gamma} \tau \rb f^+_{\xi_+ \xi_+} = 0,
\label{f+}
\end{equation}
implying
\begin{equation}
f^- = f^- \lb \xi_- + \frac{\alpha V_0}{2 c \gamma} \cos \lb \sqrt{\gamma} \tau \rb, T\rb, \quad f^+ = f^+ \lb \xi_+ - \frac{\alpha V_0}{2 c \gamma} \cos \lb \sqrt{\gamma} \tau \rb, T\rb.
\label{ff}
\end{equation}
Formulae (\ref{ff}) motivate the change of variables
\begin{equation}
\tilde \xi_- = \xi_- + \frac{\alpha V_0}{2 c \gamma} \cos \lb \sqrt{\gamma} \tau \rb, \quad \tilde \xi_+ = \xi_+ - \frac{\alpha V_0}{2 c \gamma} \cos \lb \sqrt{\gamma} \tau \rb,
\label{change}
\end{equation}
instead of $\xi_-$ and $\xi_+$, and we can now rewrite the equation for $P$ as $P_{\tilde \xi_- \tilde \xi_+} = 0$, which gives
\begin{equation}
P = g^-\lb \tilde \xi_-, \tau, T \rb + g^+ \lb \tilde \xi_+, \tau, T \rb.
\label{P}
\end{equation}
At $O \lb \varepsilon \rb$, using the averaging, we obtain
\begin{align}
g^{\pm} _{\tilde \xi_{\pm}} &= - \frac{\alpha V_0}{4 c^2 \sqrt{\gamma}} \sin \lb \sqrt{\gamma} \tau \rb f^{\pm}_{\tilde \xi_{\pm} } \pm \frac{1}{2 c} A^{\pm}\lb \tilde \xi_{\pm}, T \rb \tau \nonumber \\
&~~~~\mp \lsq  \frac{\alpha^2 V_0^2}{16 c^3 \gamma} \lb \tau - \frac{\sin \lb 2 \sqrt{\gamma} \tau \rb}{2 \sqrt{\gamma}} \rb - \frac{\alpha F_0}{2 c \sqrt{\gamma}} \sin \lb \sqrt{\gamma} \tau \rb \rsq f^{\pm}_{\tilde \xi_{\pm} \tilde \xi_{\pm}}, 
\label{g-}
\end{align}
where 
\begin{equation}
A^{\pm}  \lb  \tilde \xi_{\pm}, T \rb = \lb \mp 2 c f_{T}^{\pm} + \alpha f^{\pm} f_{\tilde \xi_{\pm}}^{\pm} + \beta c^2 f_{\tilde \xi_{\pm} \tilde \xi_{\pm} \tilde \xi_{\pm}}^{\pm} \rb_{\tilde \xi_{\pm}} - \gamma f^{\pm}.
\end{equation}
The homogeneous parts of the solutions for $g^{\pm} _{\tilde \xi_{\pm}}$ have been omitted since they just redefine the leading-order terms. To avoid secular terms, we require
\begin{equation}
\lb \mp 2 c f_{T}^{\pm} - \frac{\alpha^2 V_0^2}{8 c^2 \gamma}  f^{\pm}_{\tilde \xi_{\pm}}  + \alpha f^{\pm} f_{\tilde \xi_{\pm}}^{\pm} + \beta c^2 f_{\tilde \xi_{\pm} \tilde \xi_{\pm} \tilde \xi_{\pm}}^{\pm} \rb_{\tilde \xi_{\pm}} - \gamma f^{\pm} = 0.
\label{Ost10}
\end{equation}
Thus, we obtain two Ostrovsky equations for the left- and right-propagating waves. Equations (\ref{Ost10}) can be reduced to the standard form of the Ostrovsky equations by the change of variables
\begin{equation}
\hat \xi_{\pm} = \tilde \xi_{\pm} \mp \frac{\alpha^2 V_0^2}{16 c^3 \gamma} T.
\end{equation}
Expansion (\ref{WNLSol1}) is also substituted into the initial conditions (\ref{BousOstICnew1}), which we satisfy at the respective orders of the small parameter.

Hence, the bi-directional weakly-nonlinear solution of the Cauchy problem (\ref{BousOstOld1}), (\ref{BousOstIC1}) for the original variable $u(x, t)$ up to and including $O(\sqrt{\varepsilon})$ terms has the form \cite{KT2019}:
\begin{align}
u(x, t) &= V_{0} \frac{\sin{ ( \sqrt{\gamma} \tau ) }}{\sqrt{\varepsilon \gamma}} + F_{0} \cos{( \sqrt{\gamma} \tau )}  + f^-  + f^+ \nonumber \\ &  + \sqrt{\varepsilon} \left [ -\frac{\alpha V_0}{4 c^2 \sqrt{\gamma}} \sin \sqrt{\gamma} \tau ( f^- + f^+ )  \right . \nonumber \\
&\left . - \frac{\alpha }{2 c \sqrt{\gamma}} \left ( F_0 \sin ( \sqrt{\gamma} \tau ) + \frac{\alpha V_0^2}{16 c^2 \gamma} \sin ( 2 \sqrt{\gamma} \tau ) \right ) ( f^-_{\tilde \xi_-} - f^+_{\tilde \xi_+} )  \right ] + O(\varepsilon),
\label{WNLFV}
\end{align}
where the functions $f^{\pm} ( \tilde \xi_{\pm}, T )$ are solutions of the Ostrovsky equations 
\begin{align}
    \left( \mp 2 c f_{T}^{\pm} - \frac{\alpha^2 V_0^2}{8 c^2 \gamma}  f^{\pm}_{\tilde \xi_{\pm}}  + \alpha f^{\pm} f_{\tilde \xi_{\pm}}^{\pm} + \beta c^2 f_{\tilde \xi_{\pm} \tilde \xi_{\pm} \tilde \xi_{\pm}}^{\pm} \right)_{\tilde \xi_{\pm}} - \gamma f^{\pm} = 0,
    \label{Ost1}
\end{align}
which should be solved subject to the initial conditions
\begin{equation}
f^{\pm}|_{T=0} = \frac{1}{2}  [F ( \tilde \xi_{\pm} ) - F_0] \pm \frac{1}{2c}  \left[ \int_{-L}^{\tilde \xi_{\pm}} (V(\sigma) - V_0) \,d{\sigma} - \Bigl\langle \int_{-L}^{\tilde \xi_{\pm}} (V(\sigma) - V_0) \,d{\sigma}  \Bigr\rangle \right ] .
\end{equation}
Here $\displaystyle \tilde \xi_{\pm} = \xi_{\pm} \mp \frac{\alpha V_0}{2 c \gamma} \cos ( \sqrt{\gamma} \tau )$, i.e., there are oscillations in the phase of the waves
when $V_0 \ne 0$. 

In the reference frame moving with the linear wave speed $c$,  the Ostrovsky equations take the form
\begin{align}
   \left[ \mp f^{\pm}_T + \left(\frac{\alpha V_0 }{2 c \sqrt{\gamma \varepsilon}}\sin(\sqrt{\frac{\gamma}{\varepsilon}}T) - \frac{\alpha^2 V_0^2}{16 c^3 \gamma}  \right) f^{\pm}_{\xi_{\pm}} + \alpha_1 f^{\pm} f^{\pm}_{\xi_\pm} + \beta_1 f_{\xi_{\pm} \xi_{\pm} \xi_{\pm}}^{\pm} \right]_{\xi_{\pm}} = \gamma_1 f^{\pm},
\end{align}
where 
\begin{equation}
\alpha_1 = \frac{\alpha}{2c}, \quad \beta_1 = \frac{\beta c}{2}, \quad \gamma_1=\frac{\gamma}{2c}.
\label{coefs}
\end{equation}

\section{Cnoidal wave}
\label{sec:3}

Internal waves registered in oceanic observations often look like a cnoidal wave but with some phase and amplitude defects, which has motivated our recent study of the effect of rotation on such initial conditions in
  \cite{NTK2025}. The study was based only on the constructed weakly-nonlinear solution, and there was no comparison with direct numerical simulations for the rotation modified Miyata--Maltseva--Choi--Camassa (MMCC-$f$) parent system \cite{H2007}. Numerical modelling for the MMCC system is challenging due to the intrinsic Kelvin--Helmholtz instability. Hence, the aim of the present study is to construct and test similar solutions in the simple bi-directional setting of the BKG equation. One question of interest is stability of the uni-directional solutions with respect to counter-propagating perturbations, which are inevitably present in any numerical runs for the BKG equation.

 In what follows, we consider particular solutions of the BKG equation, which are uni-directional to leading order, and with $V_0 = 0$, leading to the following initial conditions for our numerical runs:
\begin{eqnarray}
    &&u |_{t=0} = F_0 + f^{-}|_{t=0} + O(\varepsilon),  \label{IC1}\\[1ex]
     &&{u_t}|_{t=0} =\, f^{-}_{t}|_{t=0} +O(\varepsilon). \label{IC2}
\end{eqnarray}

In this section, we perform a detailed test of the validity of the weakly-nonlinear solution for the initial conditions corresponding to a cnoidal wave. Therefore, we choose the function corresponding to  the cnoidal-wave solution of the KdV equation obtained in the limit $\gamma \to 0$ (e.g., \cite{W1974}, \cite{J1997}):
\begin{equation}
F_0 +  f^{-} =  \dfrac{6\beta_1}{\alpha_1}\Bigg\{ u_2 + (u_3 - u_2) \text{ cn}^2\Big[ \sqrt{\dfrac{u_3 -u_1}{2}}  (\xi_- - v_c T) ; m \Big] \Bigg\},
\end{equation}
where $\xi_{-} = x-ct $, $u_1 < u_2 < u_3$ are real, $v_c= 2 \beta_1 (u_1 + u_2 + u_3)$, and $m ={(u_3-u_2)}/{(u_3-u_1)}$ is the elliptic parameter.
Here, $F_0$ denotes the constant mean value of this function, and $f^{-}$ is the zero-mean part of the function.

The initial conditions for the BKG equation with $\gamma = 0$,  corresponding to the cnoidal-wave solution of the KdV equation, is given by 
\begin{eqnarray}
    u|_{t=0} &=&  \dfrac{6\beta_1}{\alpha_1}\Bigg\{ u_2 + (u_3 - u_2) \text{ cn}^2\Big[ \sqrt{\dfrac{u_3 -u_1}{2}} x; m \Big] \Bigg\} + O(\varepsilon), \\
     u_t |_{t=0} &=&  \dfrac{6 \sqrt{2} \beta_1 c}{\alpha_1} (u_3 - u_2) \sqrt{u_3 - u_1}\  \text{cn} \Big[\sqrt{\dfrac{u_3 -u_1}{2}} x; m \Big] \nonumber \\
     &\times&  \text{sn} \Big[\sqrt{\dfrac{u_3 -u_1}{2}} x; m \Big]\  \text{dn}\Big[\sqrt{\dfrac{u_3 -u_1}{2}} x; m \Big] + O(\varepsilon).
\end{eqnarray}

\begin{figure}[h]
    \centering
    \includegraphics[width=0.45\linewidth]{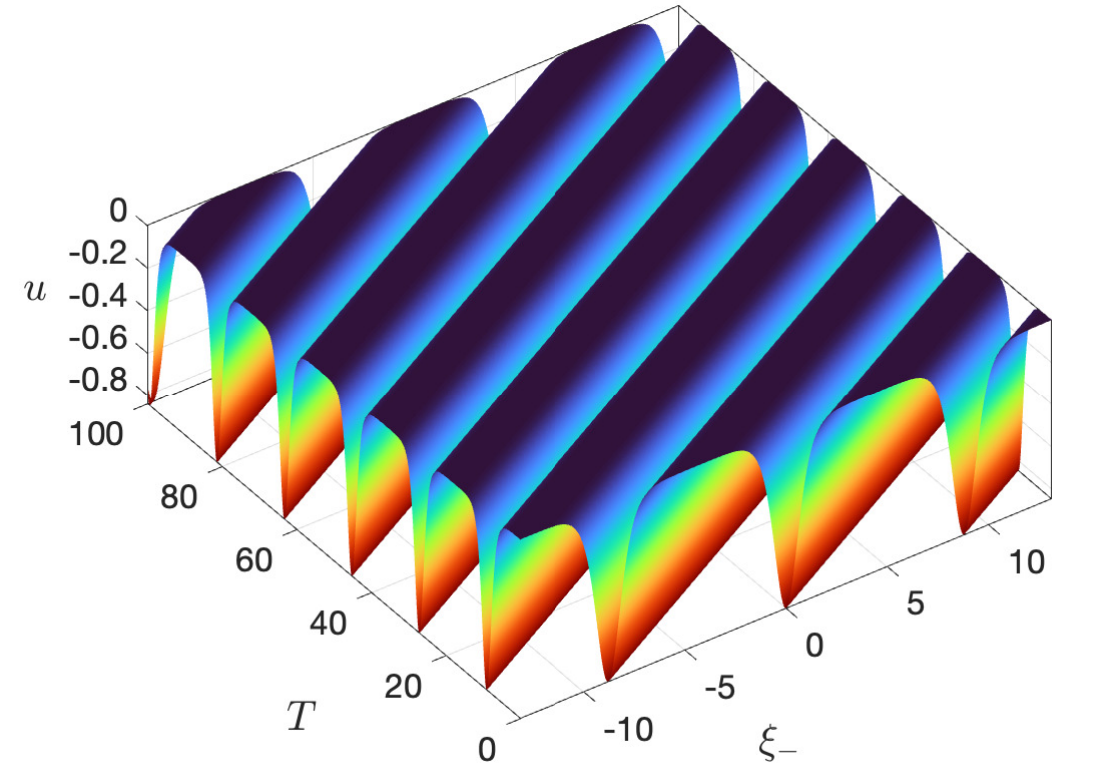}
    \includegraphics[width=0.45\linewidth]{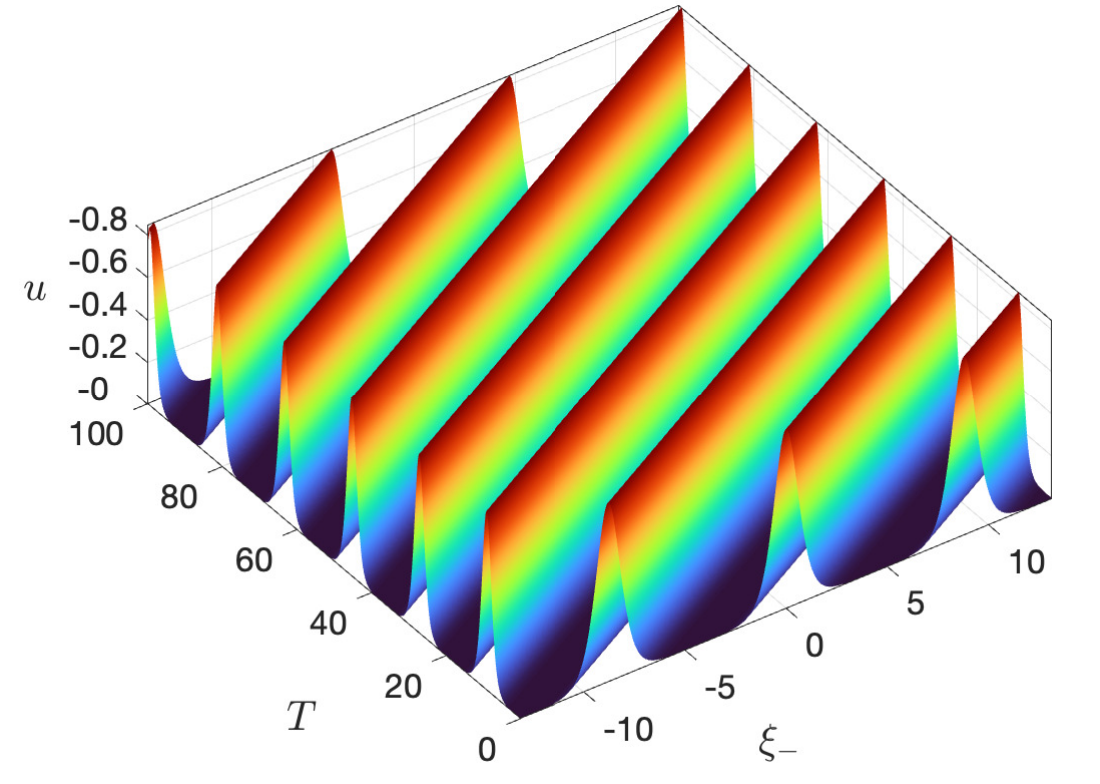}
\vspace{0.25cm}    
    \includegraphics[width=0.29\linewidth]{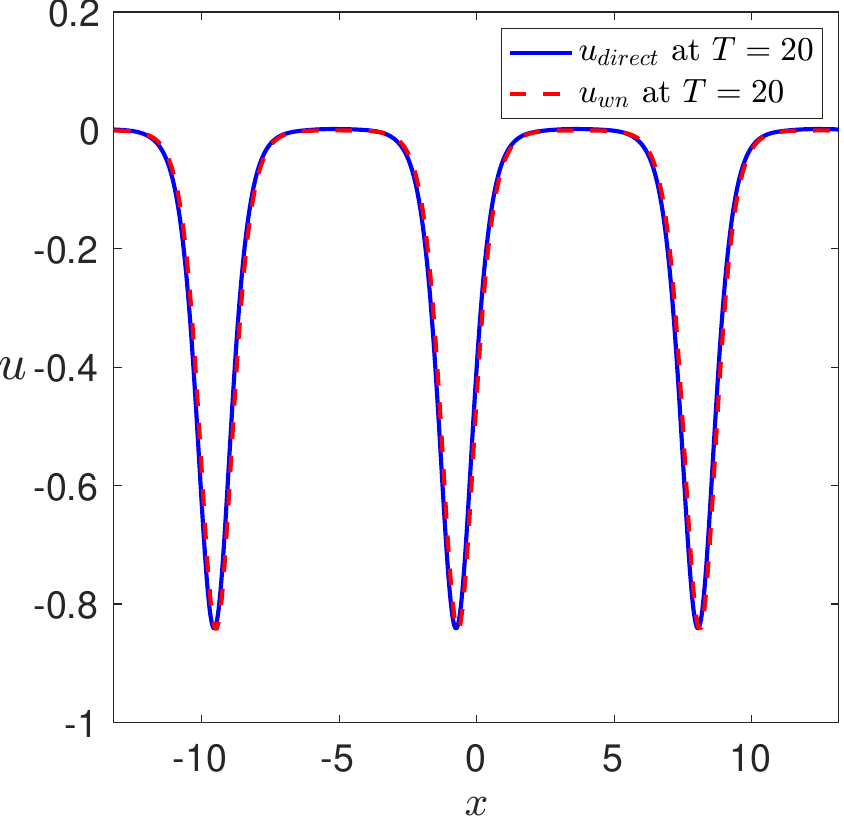} 
    \includegraphics[width=0.27\linewidth]{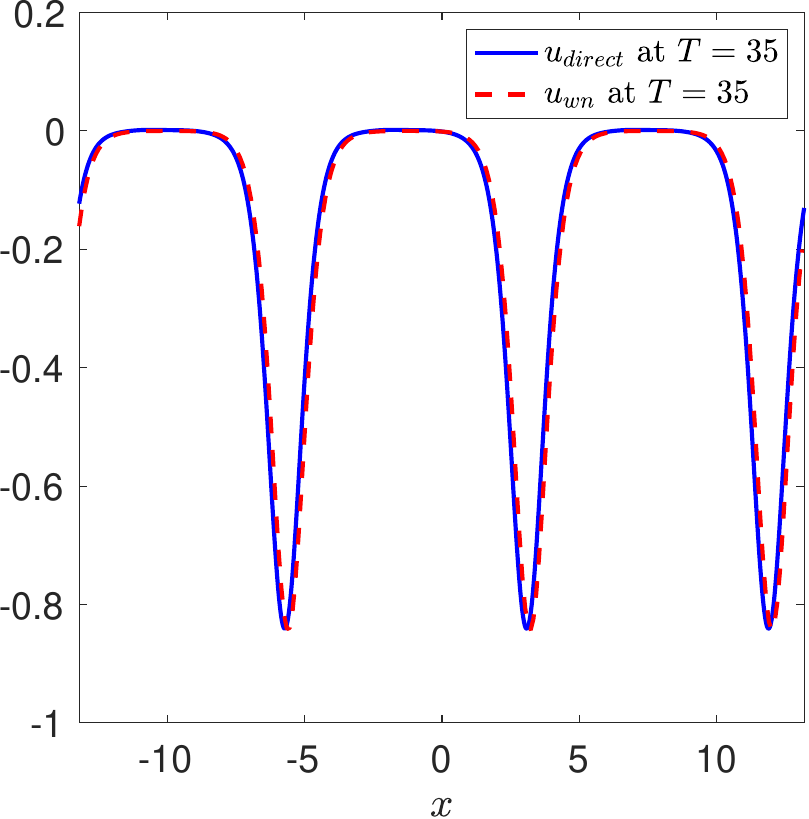}
    \includegraphics[width=0.27\linewidth]{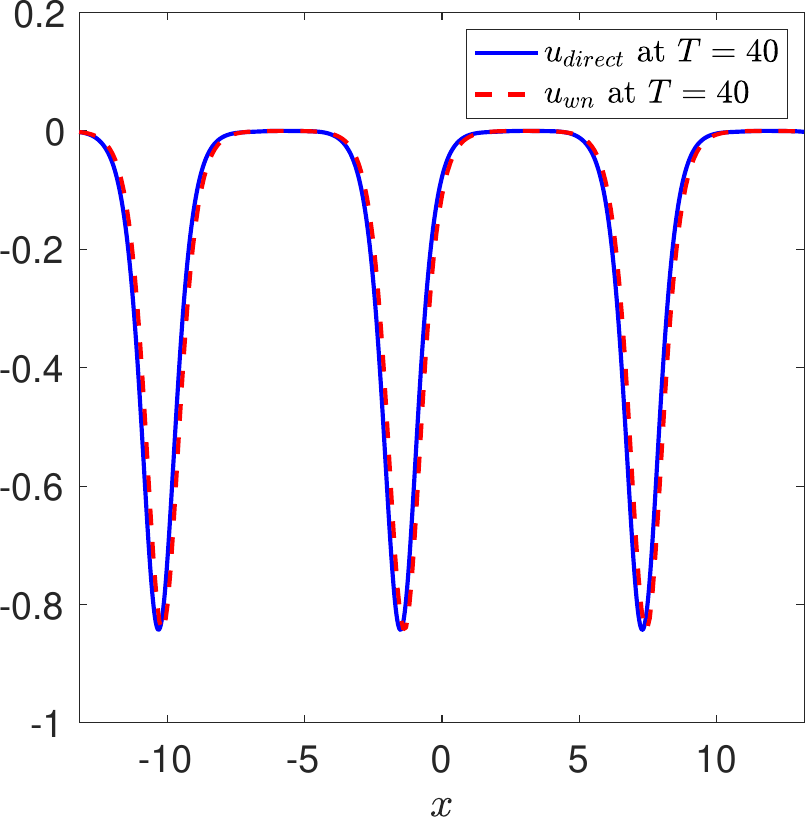} \\
\vspace{0.25cm}
    \includegraphics[width=0.29\linewidth]{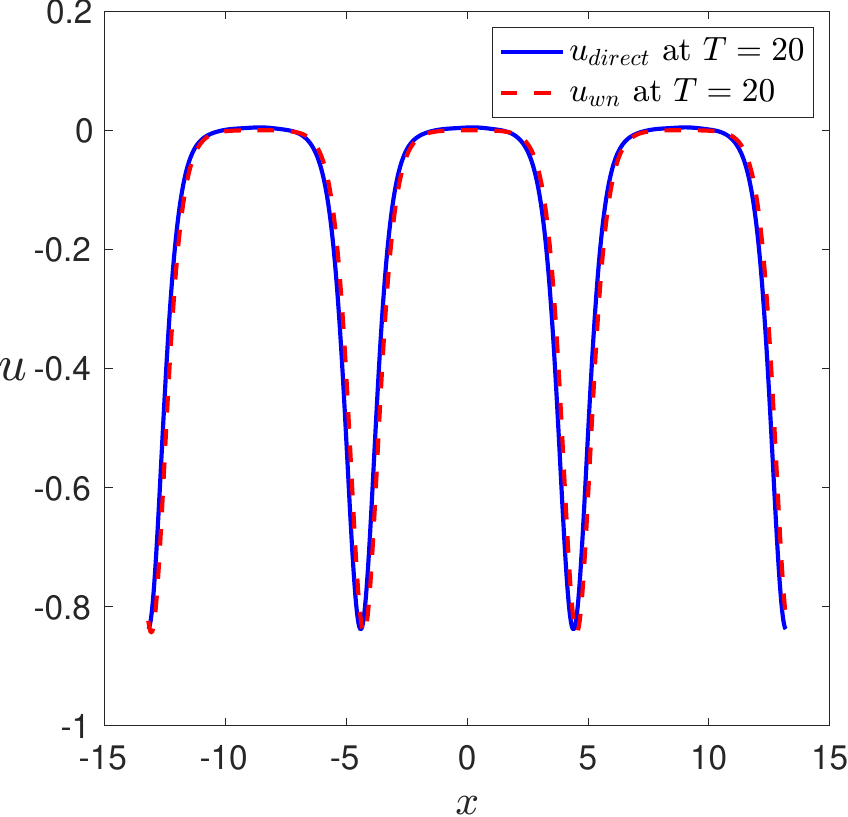} 
    \includegraphics[width=0.27\linewidth]{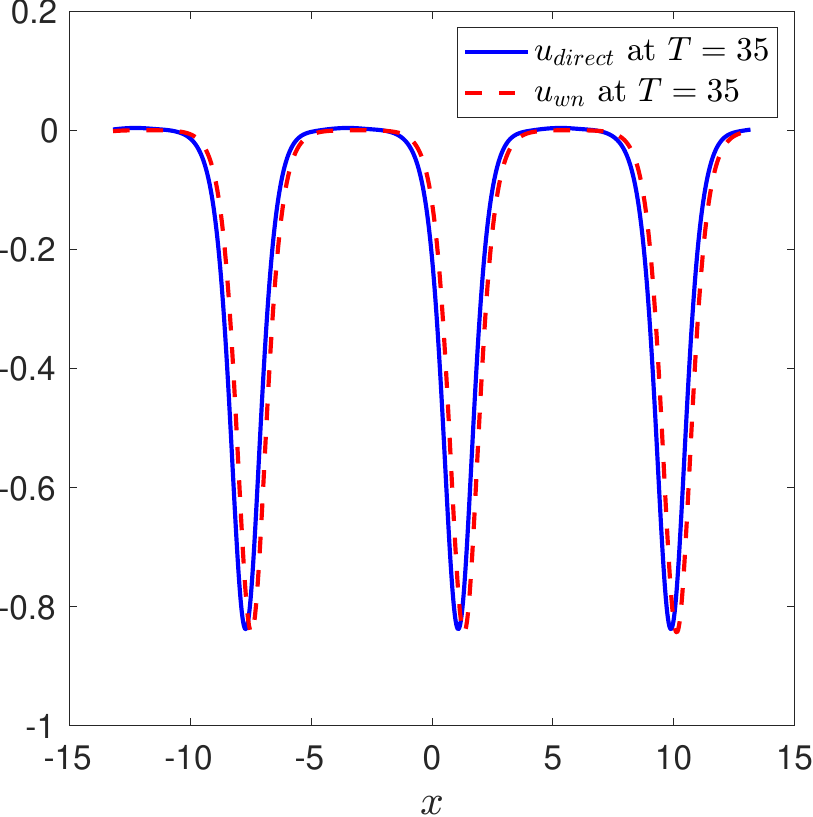}
    \includegraphics[width=0.27\linewidth]{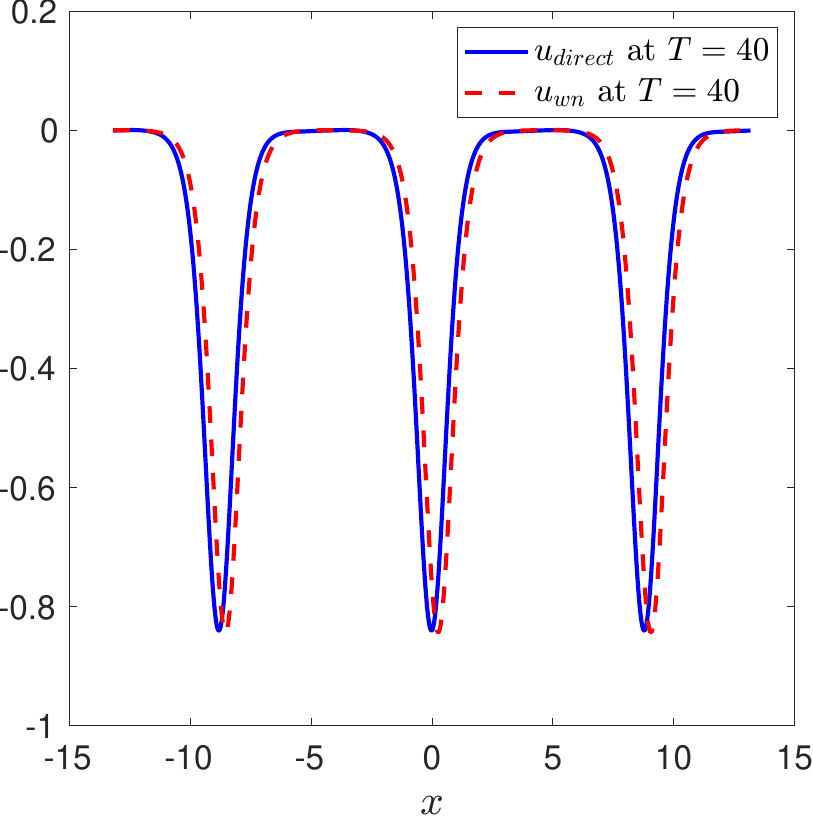}
    \caption{\footnotesize
        Evolution of the weakly nonlinear solution $u$ given by \eqref{WNLFV} for a cnoidal wave initial condition, showing 
    view from above (left) and view from below (right) (first row). Numerical parameters are $\varepsilon = 0.005,\, \alpha_1 = -1.73,\, \beta_1 =  0.08,\, \gamma_1 =0, u_1 = -10^{-3},\, u_2=0,\, u_3=3$. Comparison of the direct numerical simulations (blue, solid) and weakly nonlinear solution (red, dashed) for a cnoidal wave initial condition when $\gamma=0$ at times $T=20$ (left), $T=35$ (middle) and $T=40$ (right), computed for $\varepsilon = 0.005$ (second row) and  $\varepsilon = 0.01$ (third row).
    }
    \label{fig5.4}
\end{figure}

Figure \ref{fig5.4} shows the evolution of the weakly nonlinear solution \eqref{WNLFV} with $\gamma_1 = 0$, plotted in the moving frame $\xi_{-}$ over the total slow time $T_{\text{max}} = 100$. The computation is performed on a periodic domain of length $2L = 26.40$, using $M = 526$ Fourier modes, corresponding to a spatial step size of $\Delta x \approx 5.02 \times 10^{-2}$, and a temporal step size of $\Delta T = 5.00 \times 10^{-3}$. Numerical parameters are $\varepsilon = 0.005,\, \alpha_1 = -1.73,\, \beta_1 =  0.08,\, \gamma_1 =0, u_1 = -10^{-3},\, u_2=0,\, u_3=3$. Here,  $F_0 = -0.16$ and $V_0 = 0$. The weakly nonlinear solution is compared with the results of the direct numerical simulations for the BKG equation for $\gamma_1 = 0$  for $\varepsilon = 0.005$ and $\varepsilon = 0.01$, respectively, showing very good agreement, even in long numerical runs.

\begin{figure}
    \centering
    \includegraphics[width=0.7\linewidth]{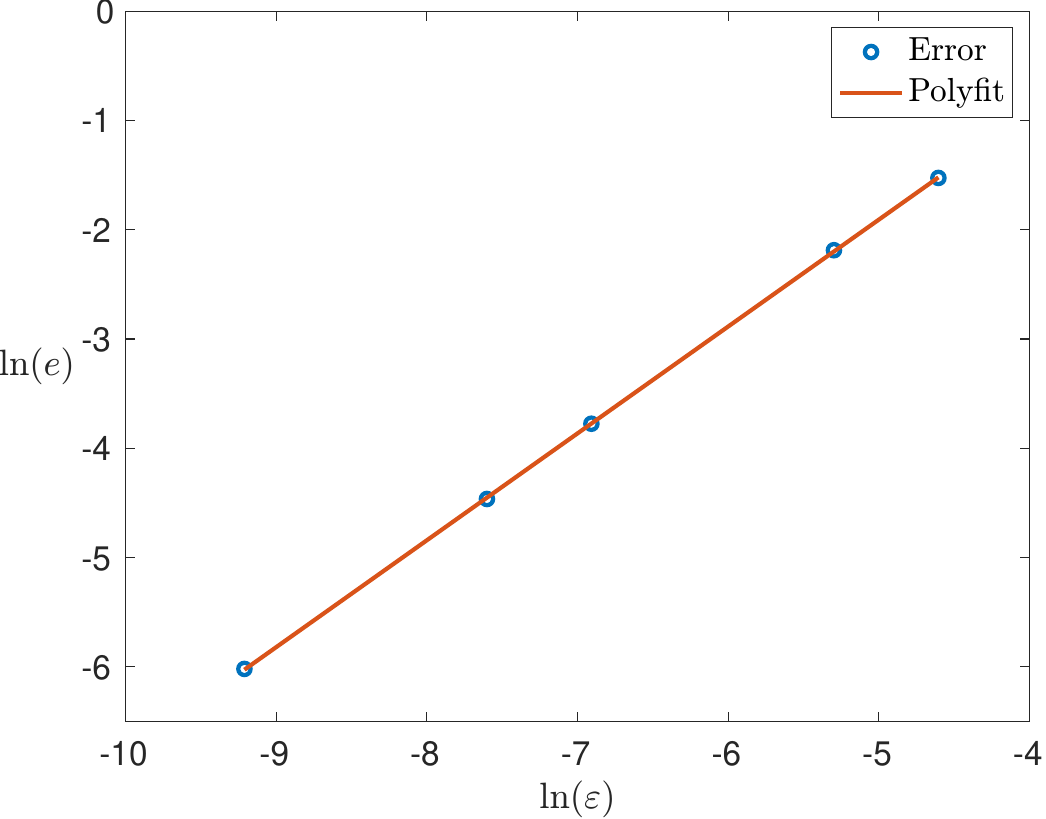}
    \caption{\footnotesize Least-squares fitting of error computed for $\varepsilon = 0.0001,\, 0.0005,\, 0.001,\, 0.005,\, 0.01$ at $T_{max} = 100$.}
    \label{fig_maxerr}
\end{figure}

Denoting the numerical solution as $u_{\text{direct}}$, and the leading-order weakly-nonlinear approximation as $u_{\text{wn}}$, we examine their agreement further by calculating the maximum pointwise absolute error in $x$ at the final moment of time, given by
\begin{equation}
e(T_{max}) = \max_{-L \leq x \leq L} \left| u_{\text{direct}}(x, T_{max}) - u_{\text{wn}}(x, T_{max}) \right|,
\label{MaxErr}
\end{equation}
at $T_{max} = 100$.
To quantify how this error scales with the small parameter $\varepsilon$, we apply a least-squares fit to a power law of the form
\begin{equation}
 e = C \varepsilon^{d},
\label{Err}
\end{equation}
where $C$ and $d$ are fitting parameters,  using the \textit{polyfit} function in MATLAB. The results are shown in
Figure \ref{fig_maxerr}, where $\ln(C) \approx 2.980$ and $d \approx 0.989$, implying that $e \sim O(\varepsilon)$, in agreement with the constructed weakly-nonlinear solution.

\begin{figure}[h]
    \centering
    \includegraphics[width=0.45\linewidth]{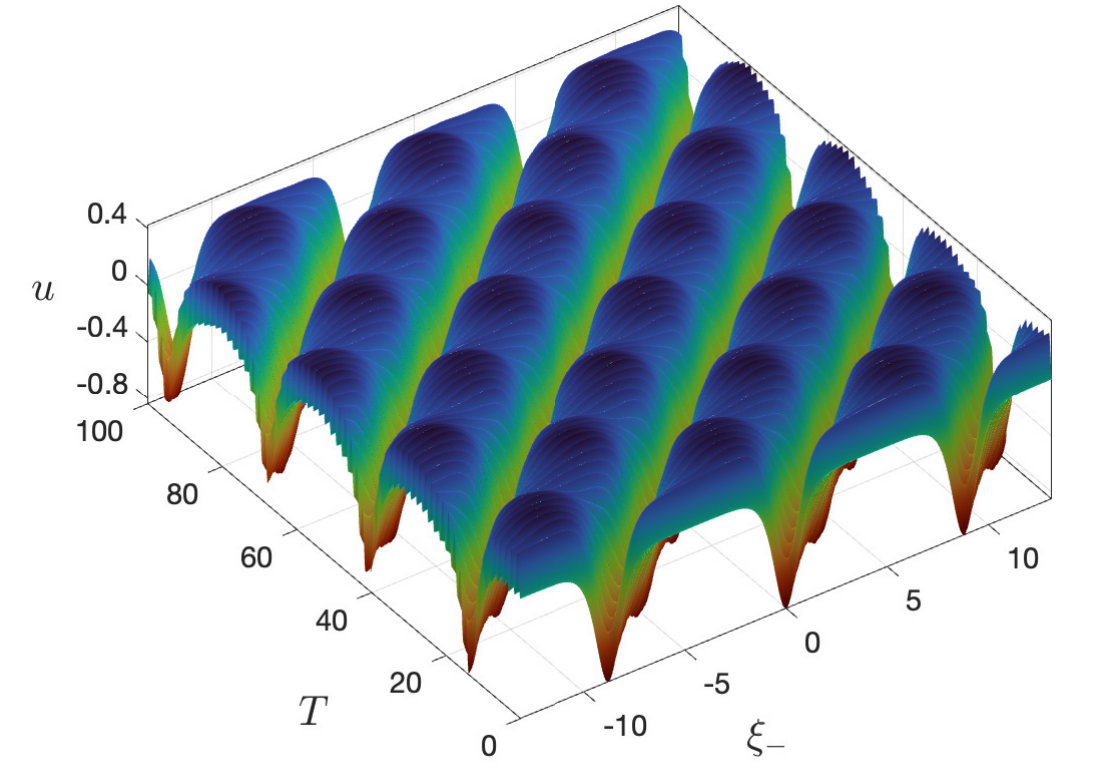}
    \includegraphics[width=0.45\linewidth]{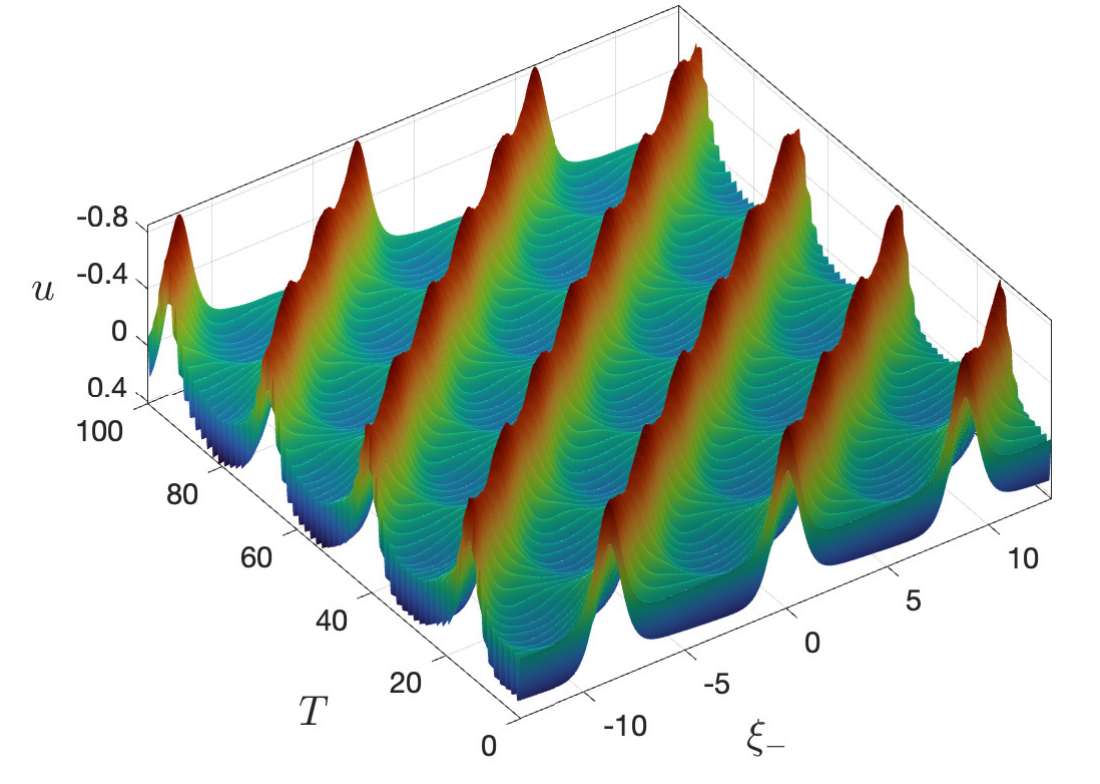}
    \vspace{0.25cm} 
    \includegraphics[width=0.29\linewidth]{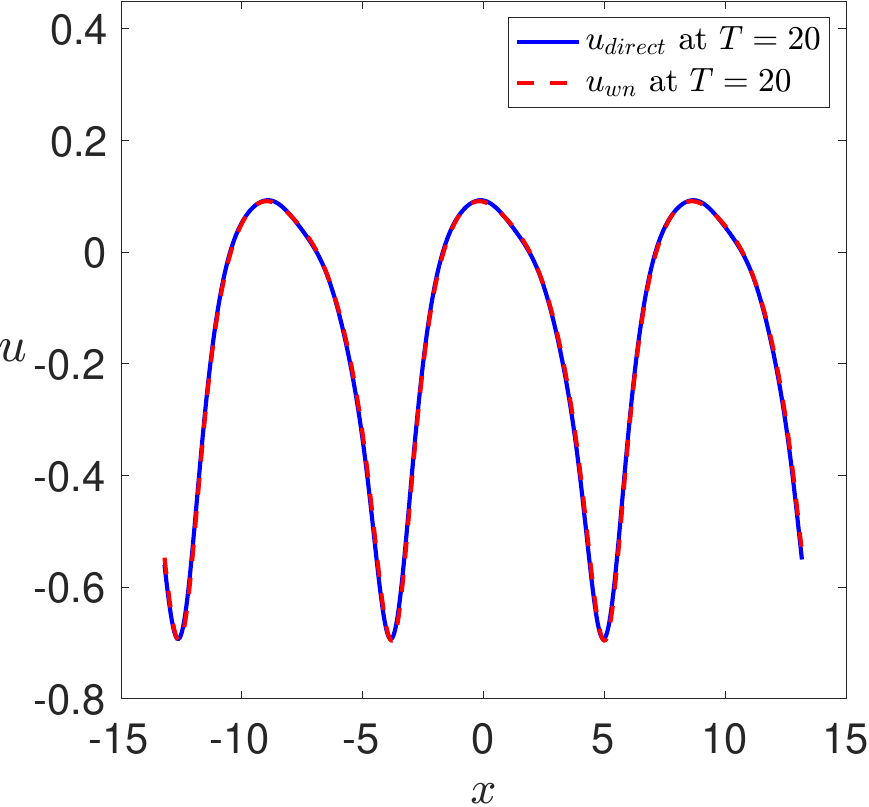} 
    \includegraphics[width=0.27\linewidth]{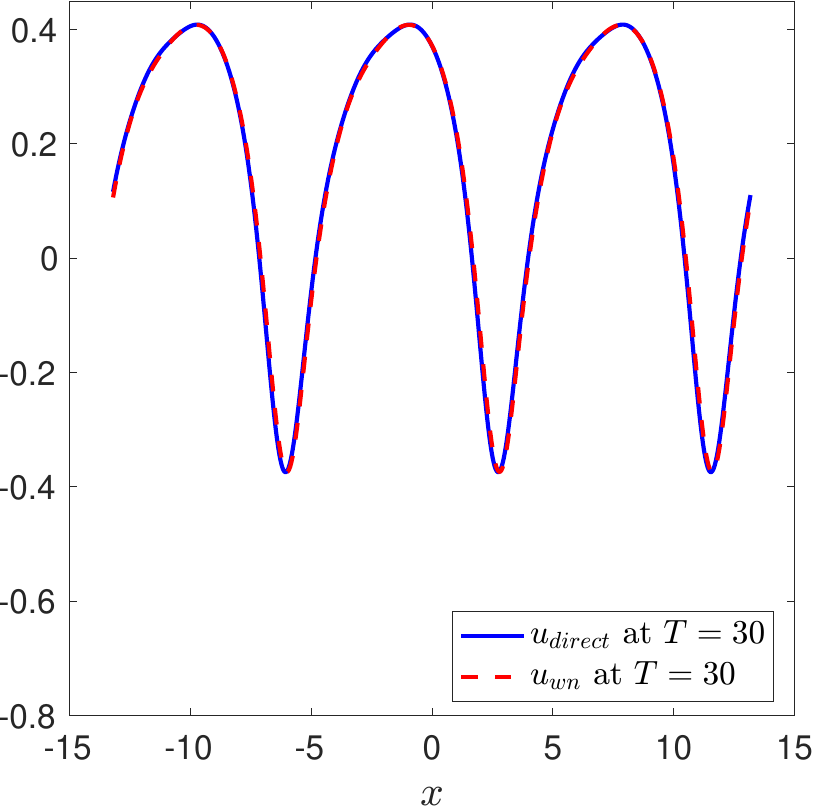}
    \includegraphics[width=0.27\linewidth]{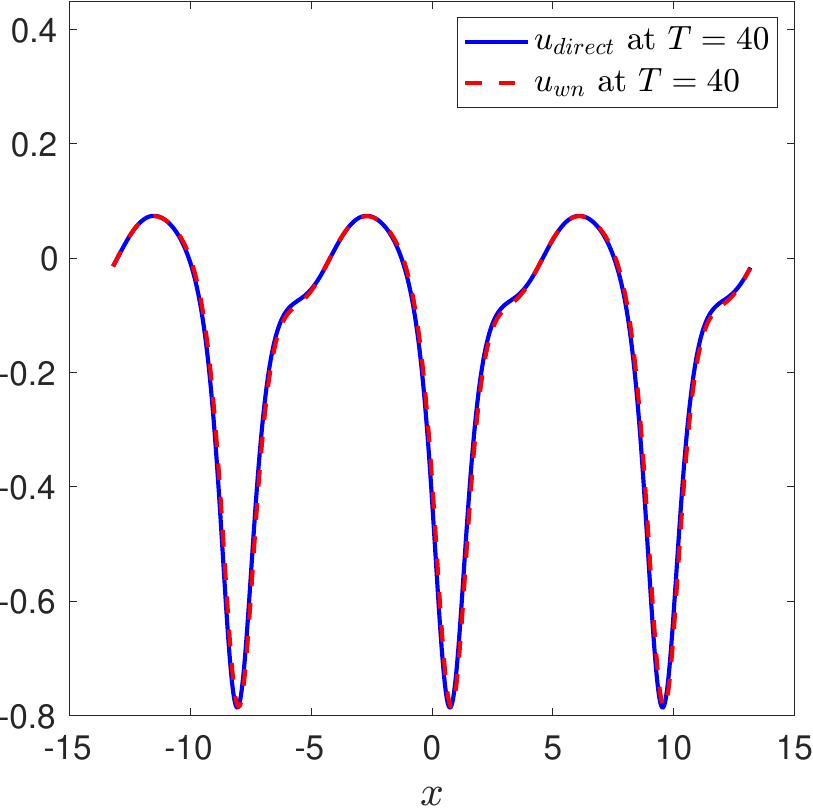}
\vspace{0.2cm}
    \includegraphics[width=0.27\linewidth]{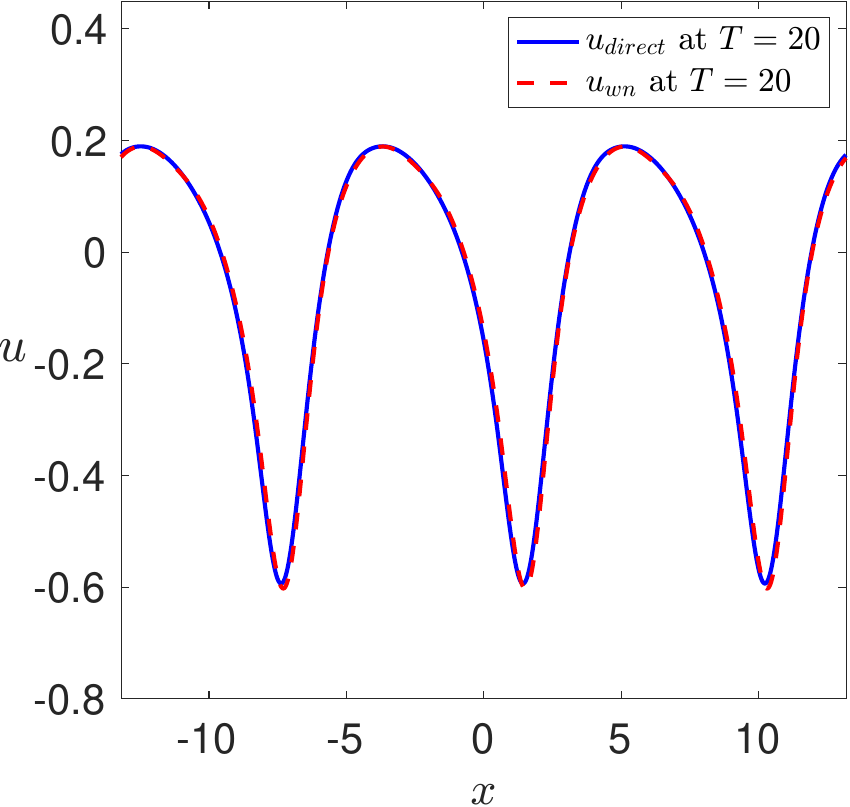} 
    \includegraphics[width=0.255\linewidth]{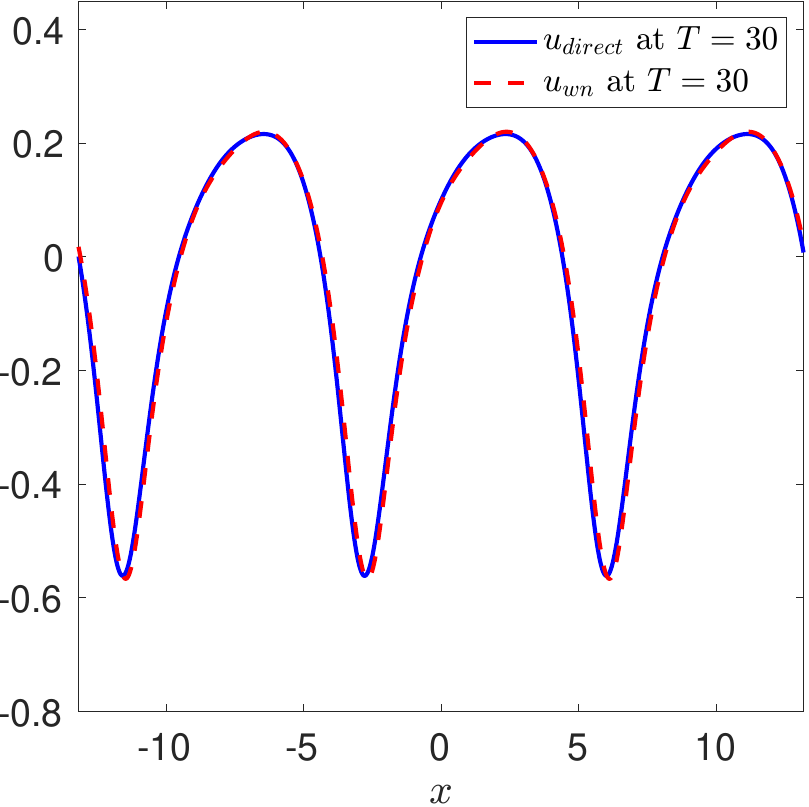}
    \includegraphics[width=0.255\linewidth]{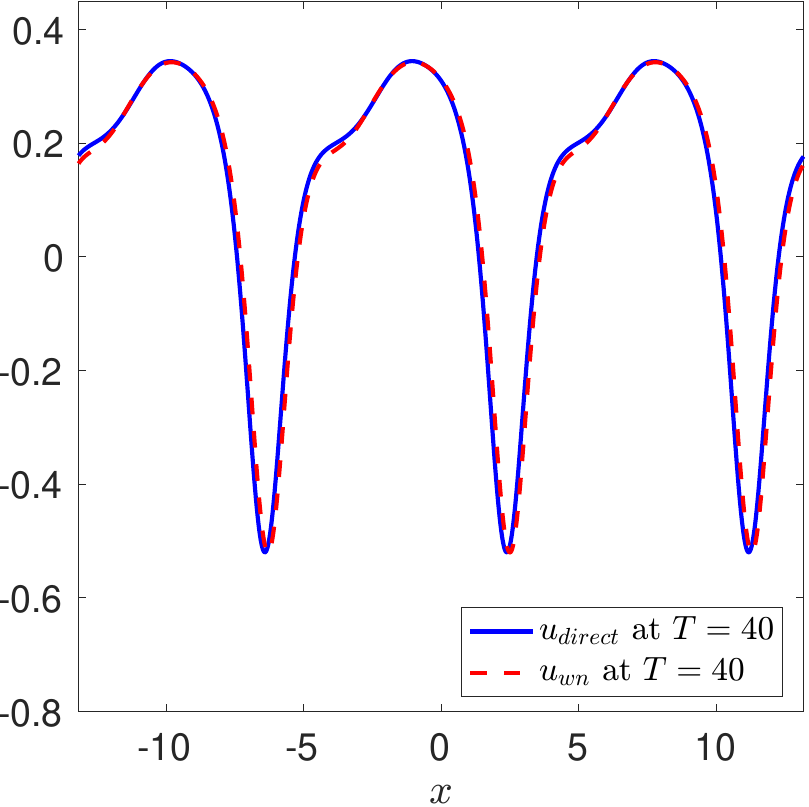}
    \caption{\footnotesize
      Evolution of the weakly nonlinear solution $u$ given by  \eqref{WNLFV} for a pure cnoidal wave initial condition with non-zero $\gamma_1$, showing
    view from above (left) and view from below (right) (first row). Numerical parameters are $\varepsilon = 0.005,\, \alpha_1 = -1.73,\, \beta_1 =  0.08,\, \gamma_1 = 0.10, u_1 = -10^{-3},\, u_2=0,\, u_3=3$.
Comparison of the profiles for a pure cnoidal wave initial condition with non-zero $\gamma_1$, on periodic boundary conditions. Direct computation (blue solid) and weakly nonlinear solution (red dash) are shown at times $T=20$ (left), $T=30$ (middle) and $T=40$ (right), computed with $\varepsilon = 0.005$ (second row) and $\varepsilon = 0.01$ (third row)..
    } 
    \label{fig5.7}
\end{figure}

Evolution of the weakly-nonlinear solution for the same cnoidal wave initial conditions, but with $\gamma_1 =0.10$ is shown in Figure \ref{fig5.7}. The numerical parameters are the same as before.
With non-zero $\gamma$, the cnoidal wave becomes modulated, with slowly oscillating amplitude and  the average value. Nevertheless, at any fixed moment of time the wave looks qualitatively similar to the original cnoidal wave, and there are no significant bursts of energy.  The weakly-nonlinear solution is compared with direct numerical simulations for the BKG equation again for $\varepsilon = 0.005$ and $\varepsilon = 0.01$. The results show that the weakly-nonlinear solution provides a good approximation, although a slight phase shift can be seen in long runs for $\varepsilon= 0.01$.  

\clearpage

\section{Cnoidal wave with an expansion defect}

It is known that localised phase defects can lead to the emergence of rogue waves \cite{HWTCH2022}.  In \cite{NTK2025} we showed that  periodicity defects introduced into the long cnoidal waves in the KdV-Ostrovsky regime can also lead to the bursts of energy, under the effect of rotation.

Motivated by observational data from \cite{SO1998},  in \cite{NTK2025} we considered cnoidal waves close to their solitonic limit and introduced  two types of periodicity defects: expansion and contraction defects, depending on whether the distance between the two neighbouring peaks was longer or shorter than the period of the wave. The expansion defect was introduced by cutting the graph at the trough and symmetrically inserting a piece of a straight line (see Figure \ref{expansion}). 
The resulting function has continuous first derivative, and {\color{black} discontinuous second derivative at two points within the domain}.  The contraction defect was introduced by symmetrically cutting away a small part close to the trough between the two neighbouring peaks, and gluing together the remaining parts of the solution. The resulting  function  has discontinuous {\color{black} first} derivative at one point within the domain, but the jump in the derivative was small because the cut was made close to an extremum. It was shown that, being smoothed in pseudospectral simulations, these functions evolved almost like travelling waves of the KdV equation. Both types of waves were long-lived, and {\color{black} cnoidal wave with an expansion defect was extremely stable, with no visible changes at the end of the long run. 

\begin{figure}
    \centering
    \includegraphics[width=0.35\linewidth]{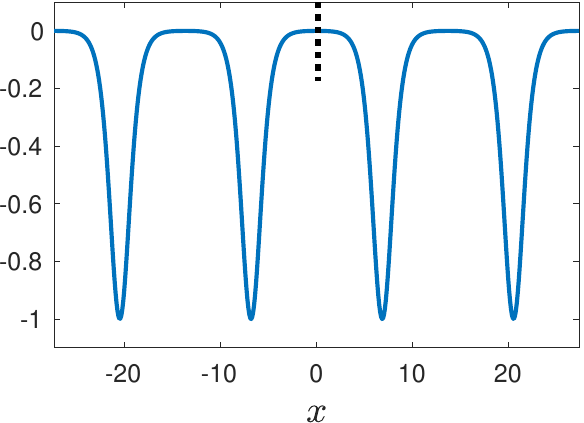}
     \includegraphics[width=0.35\linewidth]{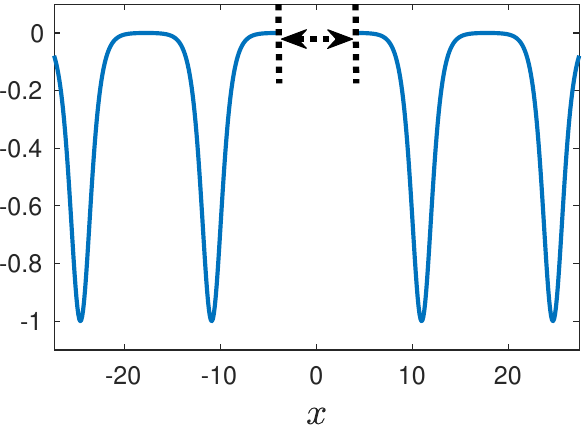}
    \caption{Schematic of construction of the generalised travelling wave of the KdV equation in the form of a cnoidal waves with an expansion defect. }
    \label{expansion}
\end{figure}

In the same paper, we proved that for both of these functions all conservation laws of the KdV equation, {\color{black} understood in the sense of a sum of integrals}, are satisfied exactly, and the cnoidal wave with an expansion defect is a broken extremal of the variational problem associated with the KdV equation. Both functions satisfy a weak formulation  \cite{NTK2025}. This approach was motivated by the studies in \cite{GNST2020,GS2022,GNS2024}, where weak solutions with singularities satisfying the Weierstrass--Erdmann corner conditions (e.g., \cite{CF1954}) were constructed  for  several equations including the Benjamin--Bona--Mahoney and conduit equations. Moreover, considering the initial conditions where a localised perturbation was superimposed on top of a cnoidal wave, we showed that both expansion and contraction defects naturally emerge in the evolution of such initial conditions in numerical simulations with periodic boundary conditions, and therefore they constitute important solutions of the KdV equation. There was no focussing of energy in these solutions in the absence of rotation, while rotation led to the emergence of a strong burst. Here, we experiment with a similar cnoidal wave with an expansion defect in a  bi-directional setting of the BKG equation.  We test the stability of this weak solution  to counter-propagating perturbations, naturally generated in these simulations because the weakly-nonlinear initial condition in not the exact solution of the full parent equation.

\begin{figure}
    \centering
    \includegraphics[width=0.45\linewidth]{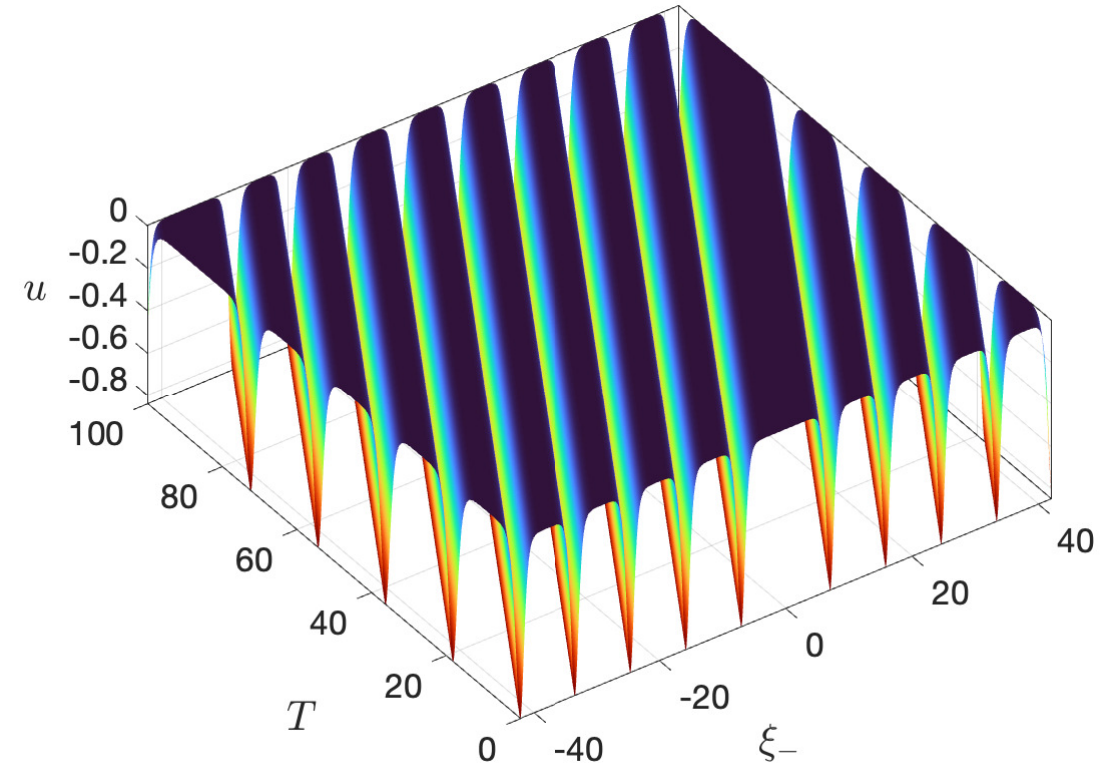}
    \includegraphics[width=0.45\linewidth]{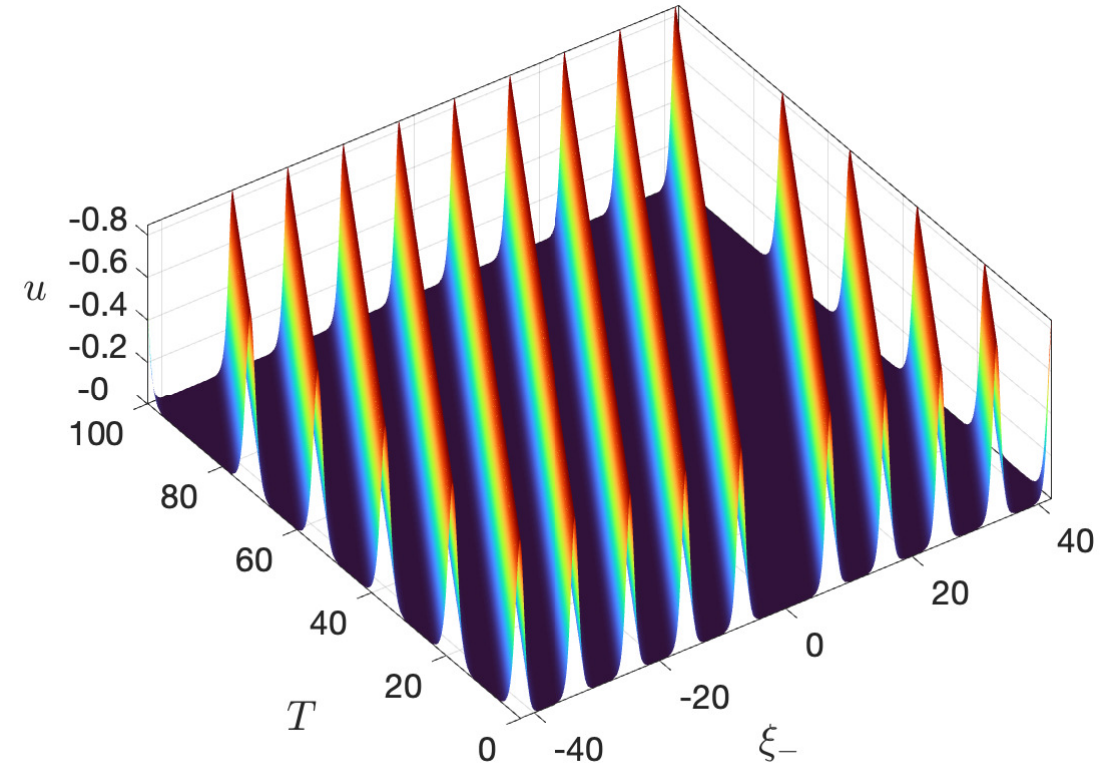} 
    \vspace{0.25cm}
    \includegraphics[width=0.29\linewidth]{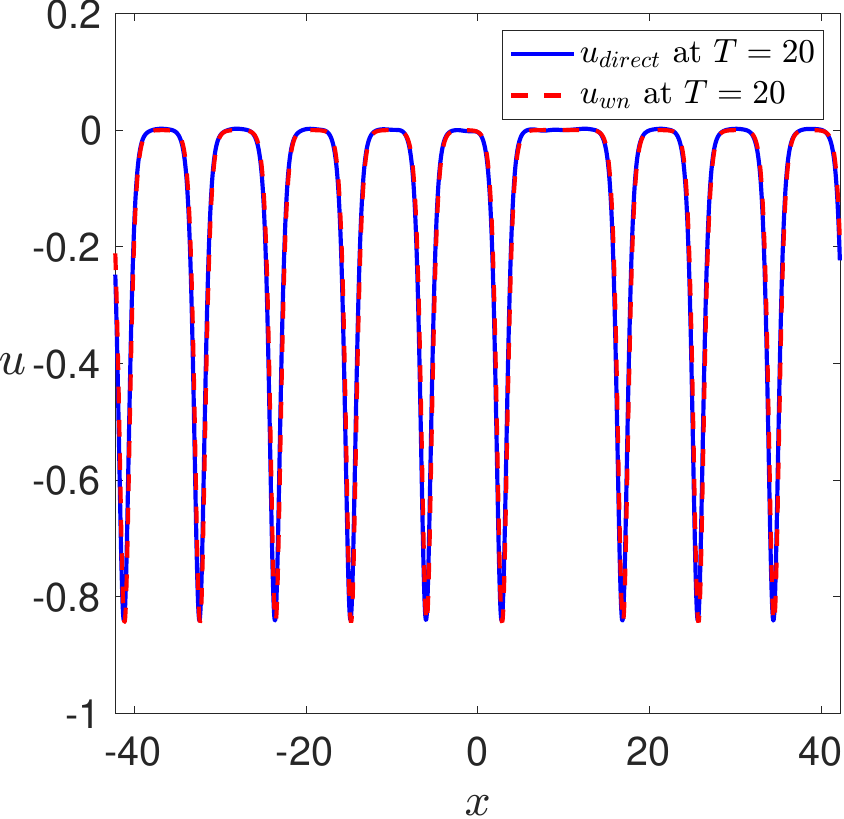} 
    \includegraphics[width=0.28\linewidth]{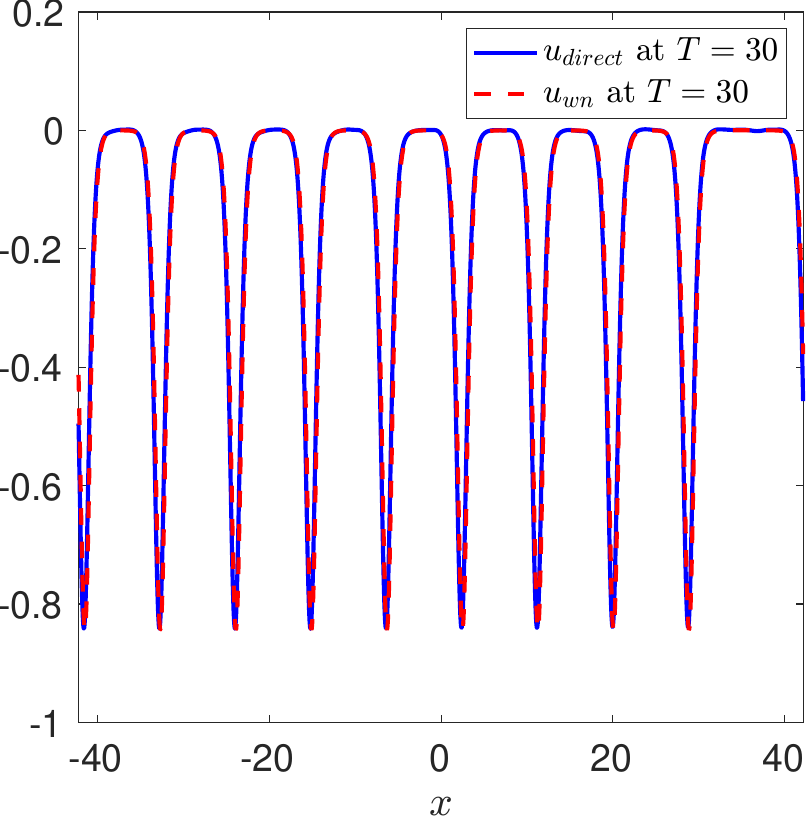}
    \includegraphics[width=0.28\linewidth]{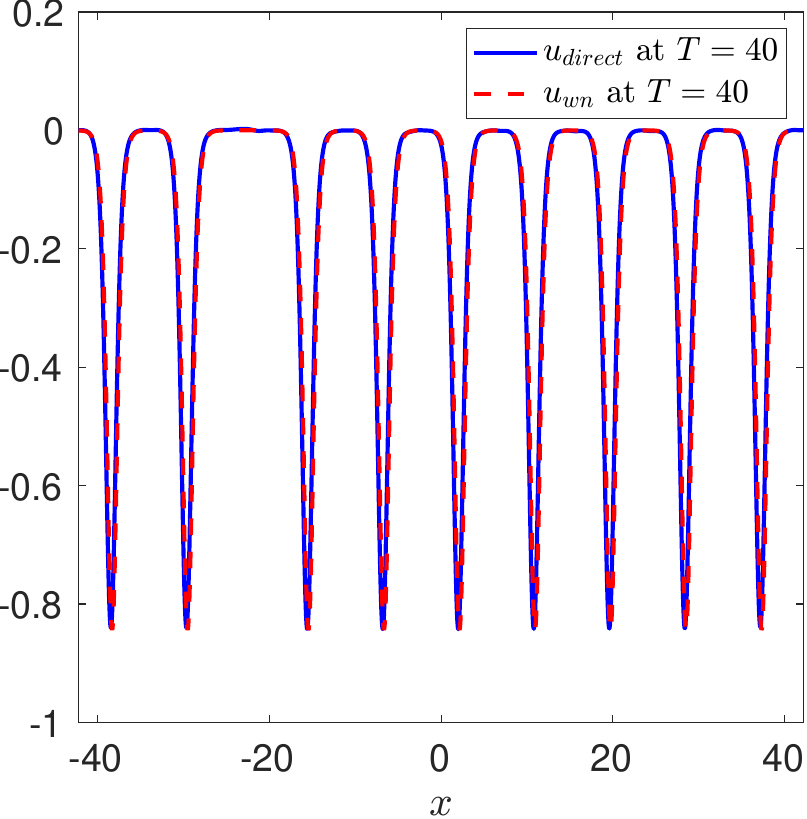}
     \vspace{0.25cm}
    \includegraphics[width=0.45\linewidth]{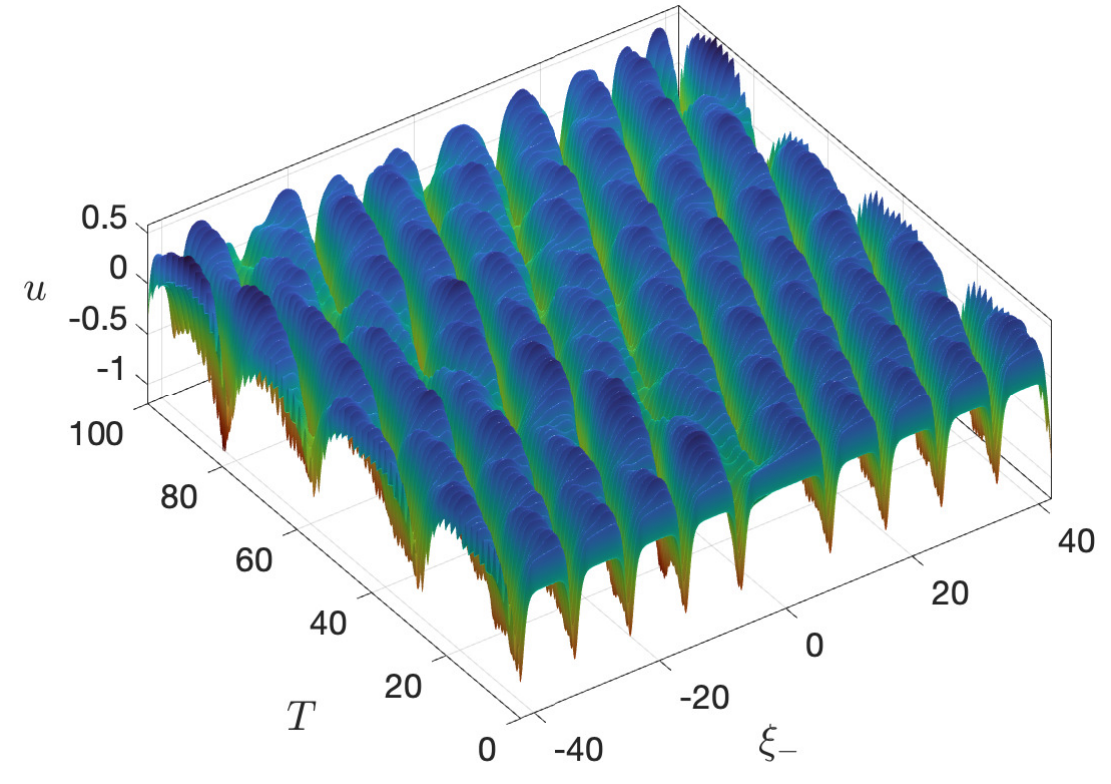}
    \includegraphics[width=0.45\linewidth]{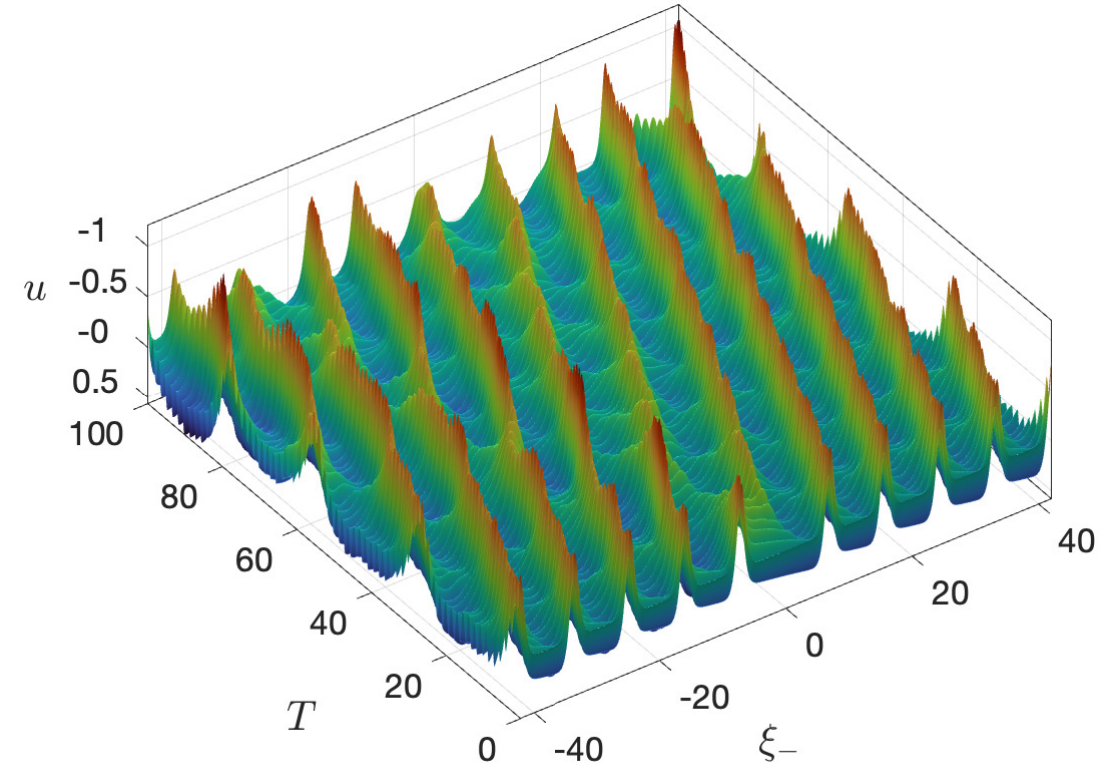}
    \includegraphics[width=0.29\linewidth]{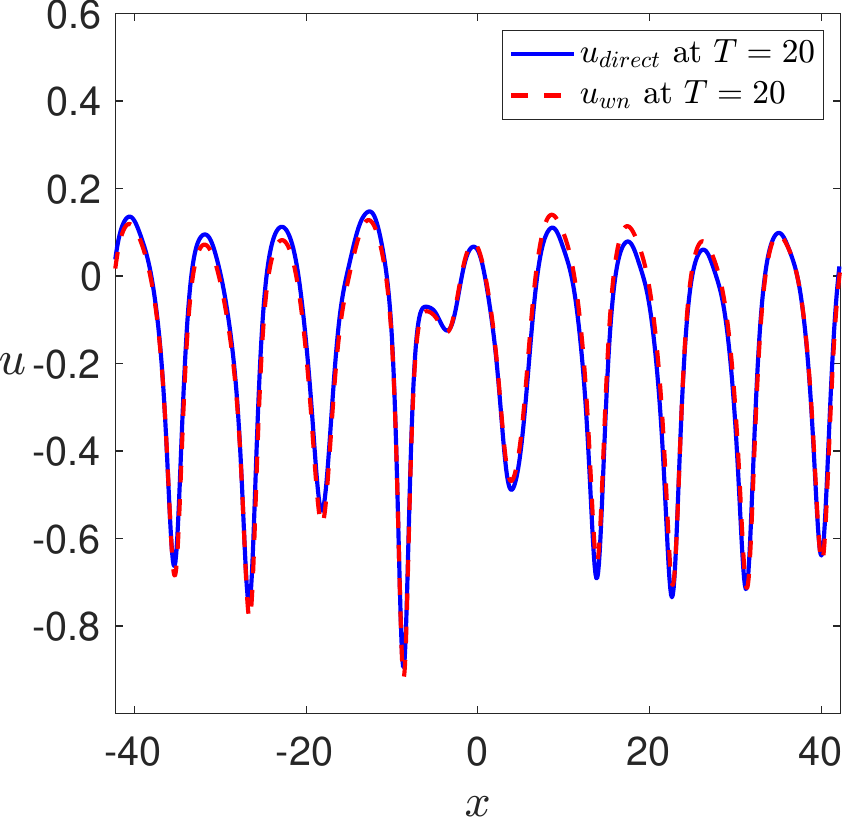} 
    \includegraphics[width=0.28\linewidth]{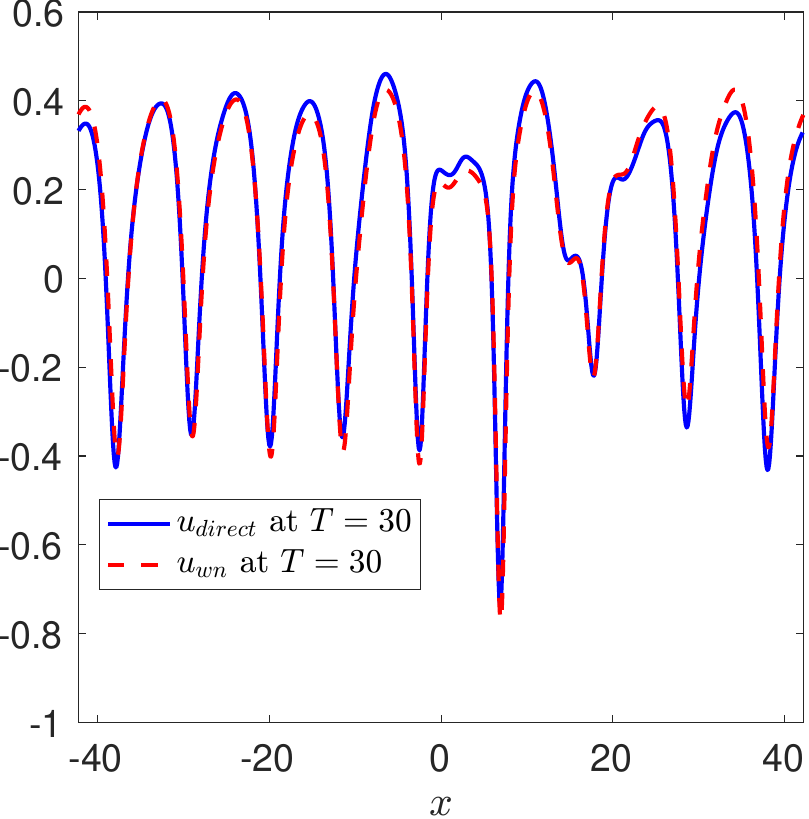}
    \includegraphics[width=0.28\linewidth]{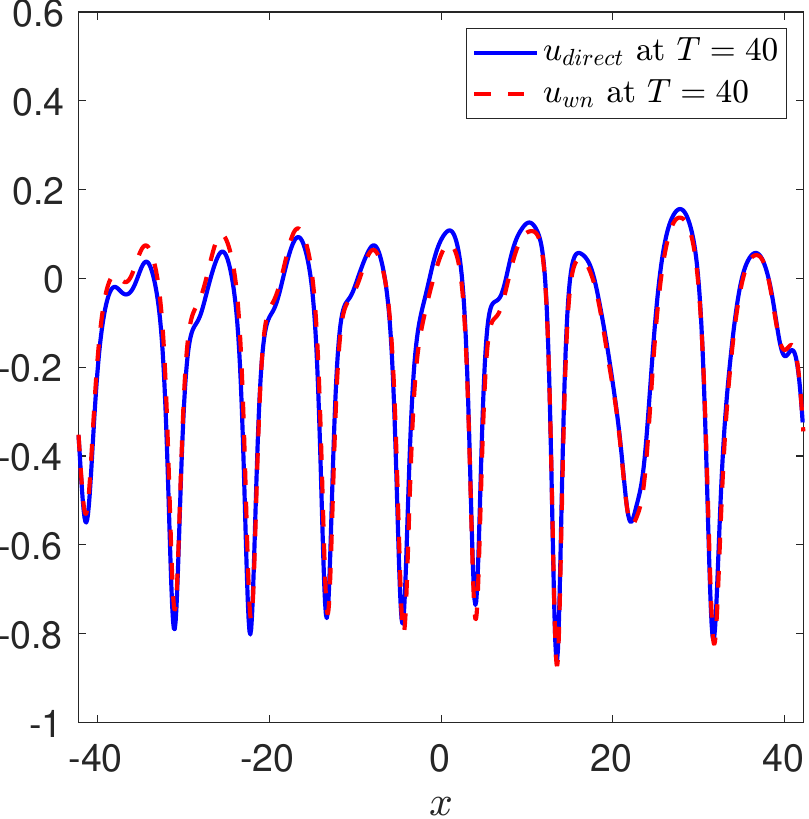}
    \caption{\footnotesize
    Evolution of the weakly nonlinear solution \eqref{WNLFV} for a cnoidal wave initial condition with expansion defect, showing view from above (left) and view from below (right) for  (a) $\gamma_1 = 0$, and (c) $\gamma_1 = 0.10$.  Comparison of the direct numerical simulations (blue, solid) and weakly nonlinear solution (red, dashed) at times $T=20$ (left), $T=30$ (middle) and $T=40$ (right) for (b) $\gamma_1 = 0$ and (d) $\gamma_1 = 0.10$, computed for $\varepsilon = 0.005$. Numerical parameters are $\varepsilon = 0.005,\, \alpha_1 = -1.73,\, \beta_1 =  0.08, u_1 = -10^{-3},\, u_2=0,\, u_3=3$.
    }
    \label{fig5.9}
\end{figure}

\clearpage
The evolution of  the cnoidal wave with an expansion defect  is shown in  Figure \ref{fig5.9}, for $\gamma_1 = 0$ ( first and second rows) and $\gamma_1 = 0.10$ (third and fourth rows). The simulation parameters used are: $\varepsilon = 0.005$, $2L = 84.48$, number of modes $M = 1688$, spatial step $\Delta x \approx 5.00 \times 10^{-2}$, and temporal step $\Delta T = 5.00 \times 10^{-3}$, with a total simulation time of $T_{\text{max}} = 100$. Despite the presence of the counter-propagating simulations, the evolution of this initial condition is similar to that reported in \cite{NTK2025}: the cnoidal wave with an expansion defect is very stable and close to a travelling wave when $\gamma_1 = 0$, while non-zero $\gamma_1$ leads to the emergence of a strong left-propagating burst of energy. 

Further, in Figure \ref{expansion_zoom} we show a close-up view of the evolution of a cnoidal wave with an expansion defect for $\gamma_1 = 0$ (KdV evolution)  for two cases: a short defect and a long defect. In both cases, there is no visible evolution, and the amplitude variations at the end of long pseudospectral runs are of order $10^{-4}$, which is illustrated in the right column.

\begin{figure}
	\centering
    \includegraphics[width=0.4\linewidth]{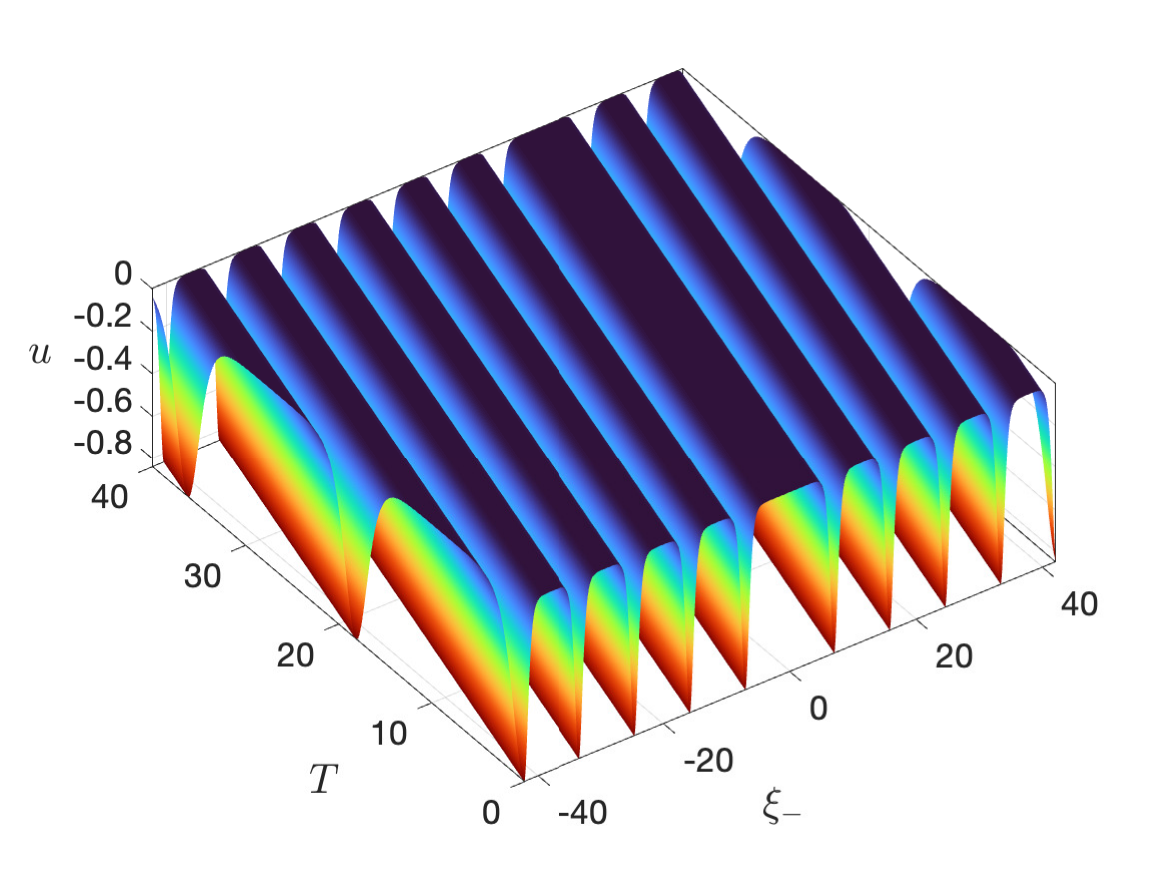}
    \includegraphics[width=0.4\linewidth]{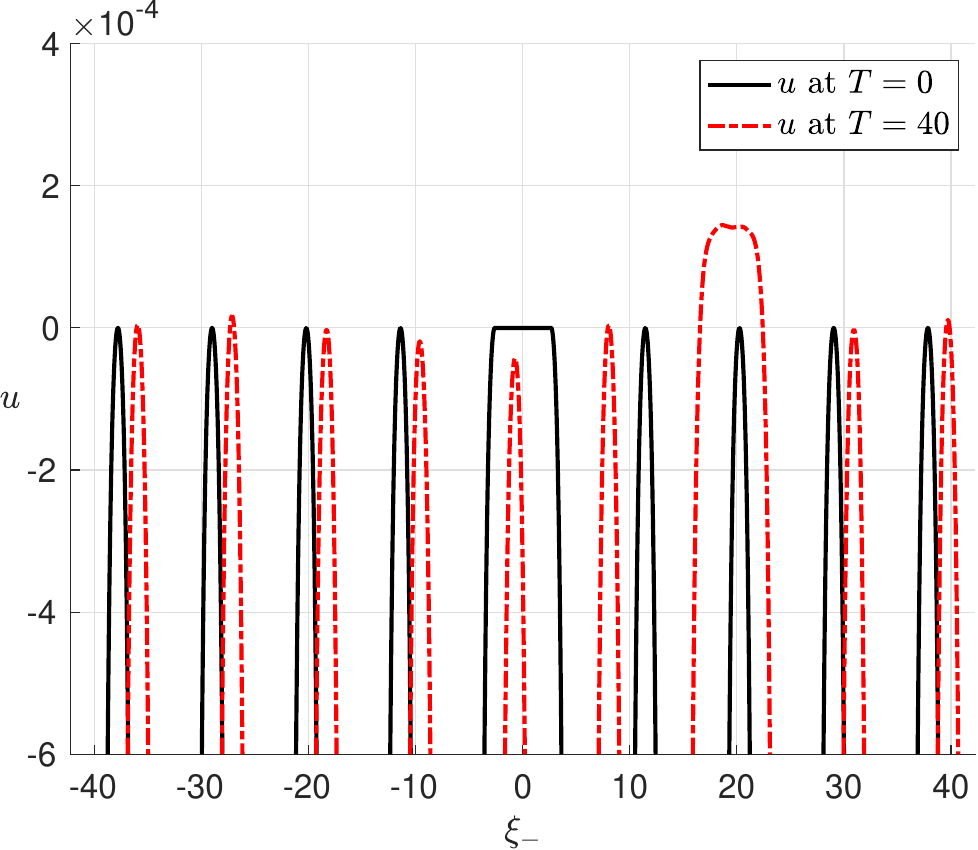}\\[1ex]
    
    \includegraphics[width=0.4\linewidth]{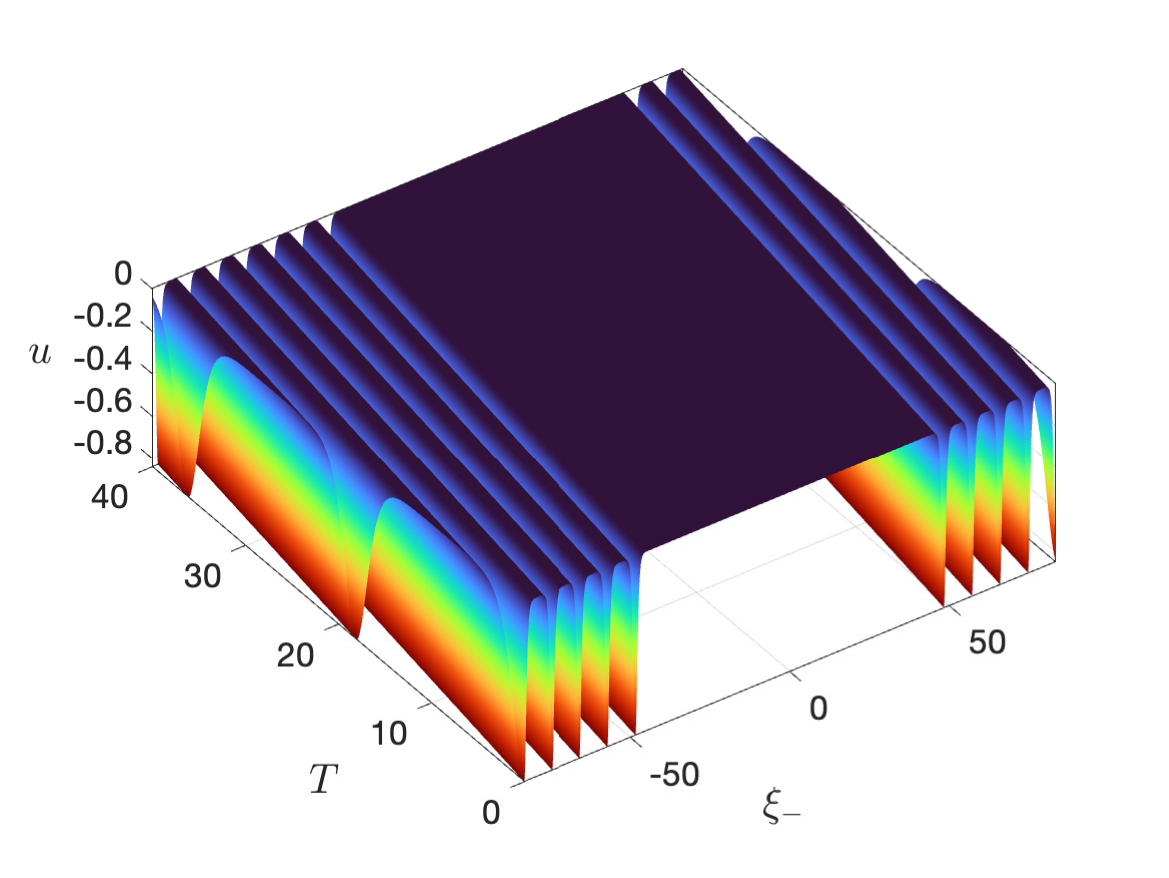}
    \includegraphics[width=0.4\linewidth]{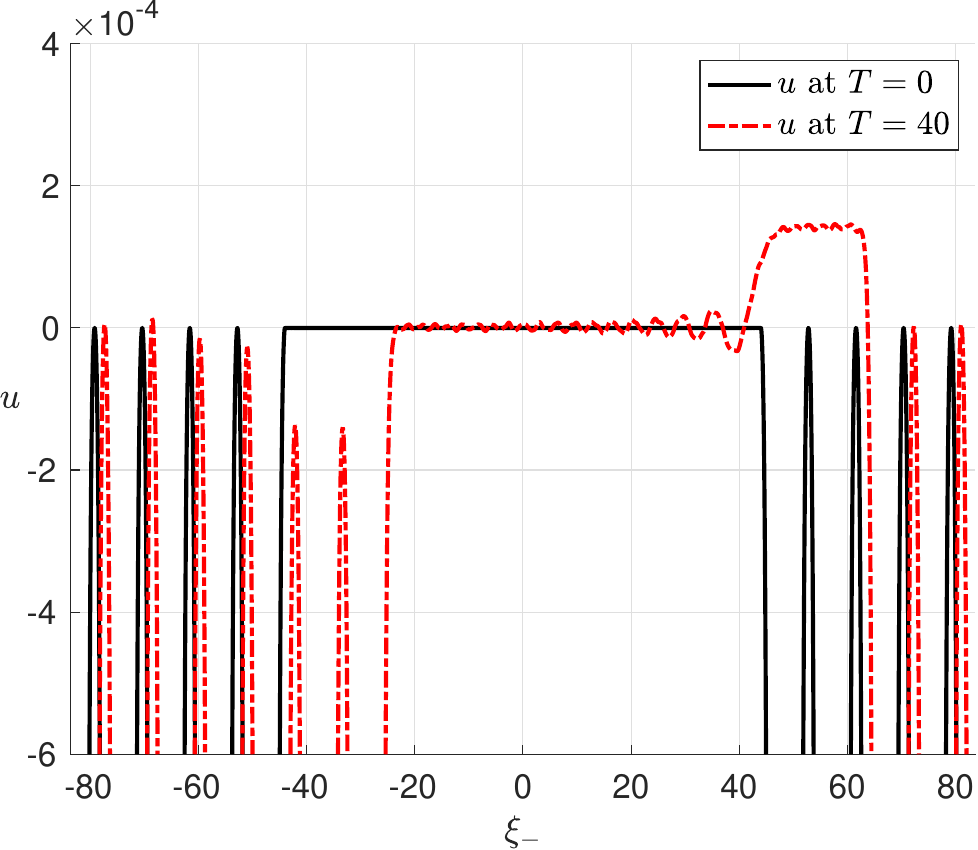}
    \caption{Evolution of a cnoidal wave with an expansion defect of length $1.2 \tilde \lambda$  (first row) and $20 \tilde \lambda$ (second row), where $\tilde \lambda$ is  the wavelength of the cnoidal wave, computed for the elliptic modulus $m\approx 0.9997$. Numerical parameters are $\varepsilon = 0.005,\, \alpha_1 = -1.73,\, \beta_1 =  0.08, \gamma_1 = 0, u_1 = -10^{-3},\, u_2=0,\, u_3=3$. }
    \label{expansion_zoom}
\end{figure}

\clearpage

\section{Cnoidal wave with a localised perturbation}

In this section we initiate numerical runs for the BKG equation with the initial conditions in the form of  a KdV cnoidal wave with a generic localised perturbation superimposed on top of the cnoidal wave: 
\begin{align}
    u|_{t=0} = \dfrac{6\beta}{\alpha}\Bigg\{ u_2 + (u_3 - u_2) \text{ cn}^2\Big[ (\xi_{-}  - v_c T) \sqrt{\dfrac{u_3 -u_1}{2}} ; m \Big] \Bigg\}_{t=0} \nonumber  \\
    + A_1 \text{sech}^2 \Big[ A_2 (x -\xi_0)  \Big]. 
    \label{local}
\end{align}
 Here, as before, $u_1 < u_2 < u_3$ are real, $v_c = 2 \beta_1 (u_1 + u_2 + u_3)$. 
$m ={(u_3-u_2)}/{(u_3-u_1)}$, and $A_1$, $A_2$, $\xi_0$ are arbitrary constants.  

The solution is computed over a larger spatial domain to accommodate nine wave peaks for $\gamma_1 = 0$ and $\gamma_1 = 0.10$. The computational domain is $2L = 84.48$, number of modes $M = 1688$, spatial step $\Delta x \approx 5.00 \times 10^{-2}$, and temporal step $\Delta T = 5.00 \times 10^{-3}$, with a total simulation time of $T_{\text{max}} = 100$. The results are shown in Figure \ref{fig5.10} for $\varepsilon = 0.005,\, \alpha_1 = -1.73,\, \beta_1 =  0.08, \, u_1 = -10^{-3},\, u_2=0,\, u_3=3, A_1 = -0.8, \, A_2 = 1, \, \xi_0 = 1.3$.

When $\gamma_1 = 0$, the initial condition generates a fast bright breather on a cnoidal wave and a slow dark breather on a cnoidal wave, which can be described analytically \cite{KM1975, HMP2023}, and a clearly identifiable expansion defect. All generated components are stable in the long pseudospectral run. When $\gamma_1 = 0.10$, the initial condition generates several bursts of energy, propagating to the left (in this moving reference frame), since all individual components of the evolution for $\gamma_1 = 0$ result in the generation of such bursts (see \cite{NTK2025} for the details). Comparisons between the constructed weakly-nonlinear solutions and the direct numerical simulations for the BKG equation are also presented in the same figure, showing good agreement for both cases, $\gamma_1 = 0$ and $\gamma_1 = 0.10$.

Next, in Figure \ref{fig5.13} we present similar results for another set of parameters given by $\varepsilon = 0.005,\, \alpha_1 = 2.25,\, \beta_1 =  0.06, \, u_1 = -10^{-3},\, u_2=0,\, u_3=3,  A_1 = 0.8,\,  A_2 = 1, \, \xi_0 = 1.3$, while $\gamma_1 = 0$ or $0.10$, as before. Here, the nonlinearity coefficient  $\alpha_1$ is positive, and the waves are waves of elevation. When $\gamma_1 = 0$, the initial condition again generates a fast bright 
and a slow dark breather on a cnoidal wave, as well as both an expansion and contraction defects, next to each other. All generated wave components were stable in the long pseudospectral runs, and in presence of the counter-propagating perturbations in the parent system. The effect of nonzero $\gamma_1$ was similar to the first case, with all initial wave components generating several left-propagating bursts. The comparison between the weakly-nonlinear solution and direct simulations for the BKG equation again showed good agreement. 

The simulations clearly illustrate the significant role of the generalised travelling waves of the KdV equation in the form of cnoidal waves with an expansion / contraction defects in the solutions of such Cauchy problems. These solutions are generated in order to maintain the periodic boundary conditions during the complicated evolution involving generation of bright and dark breathers on the cnoidal wave background. 

Finally, we note that such initial conditions can lead to the generation of a rogue wave, shown in Figure \ref{fig5.17} for the same main parameters as in Figure \ref{fig5.13}, but with $A_1 = 0.6,\, A_2 = 0.6,\, \xi_0 = 2.6$. The rogue wave is generated around $T = 7.75$, and it slowly and persistently propagates to the left, disappearing and reappearing again at a later time.

\begin{figure}
    \centering
    \includegraphics[width=0.45\linewidth]{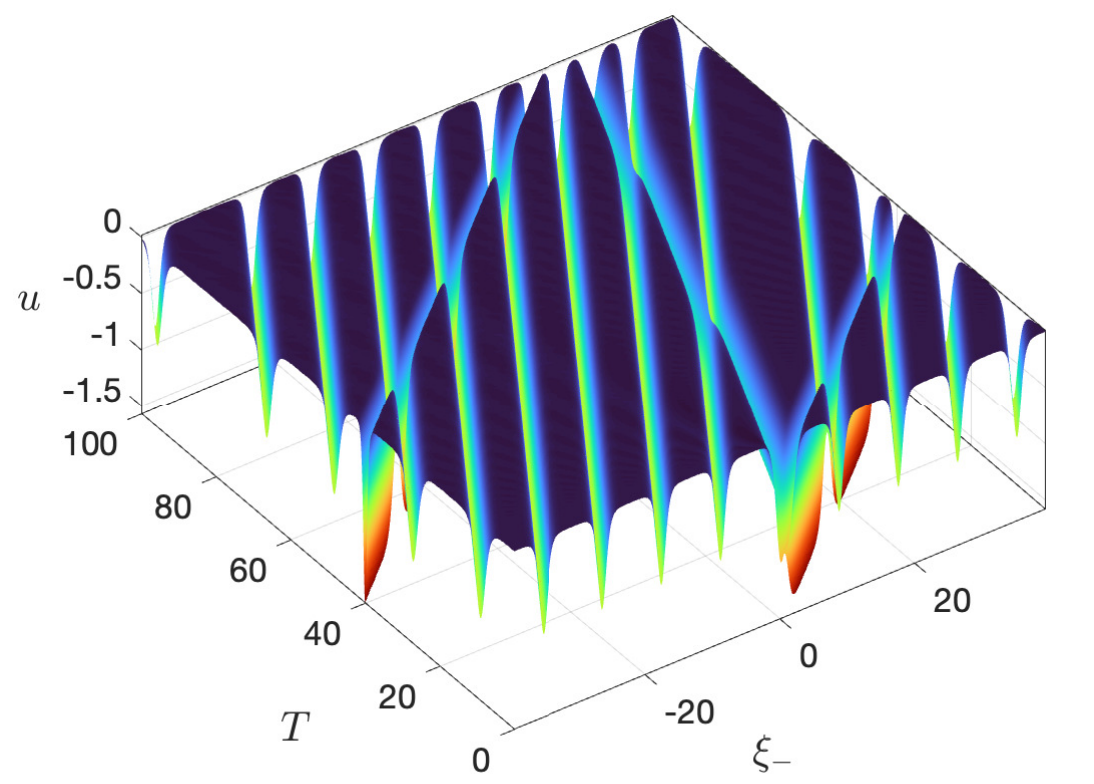}
    \includegraphics[width=0.45\linewidth]{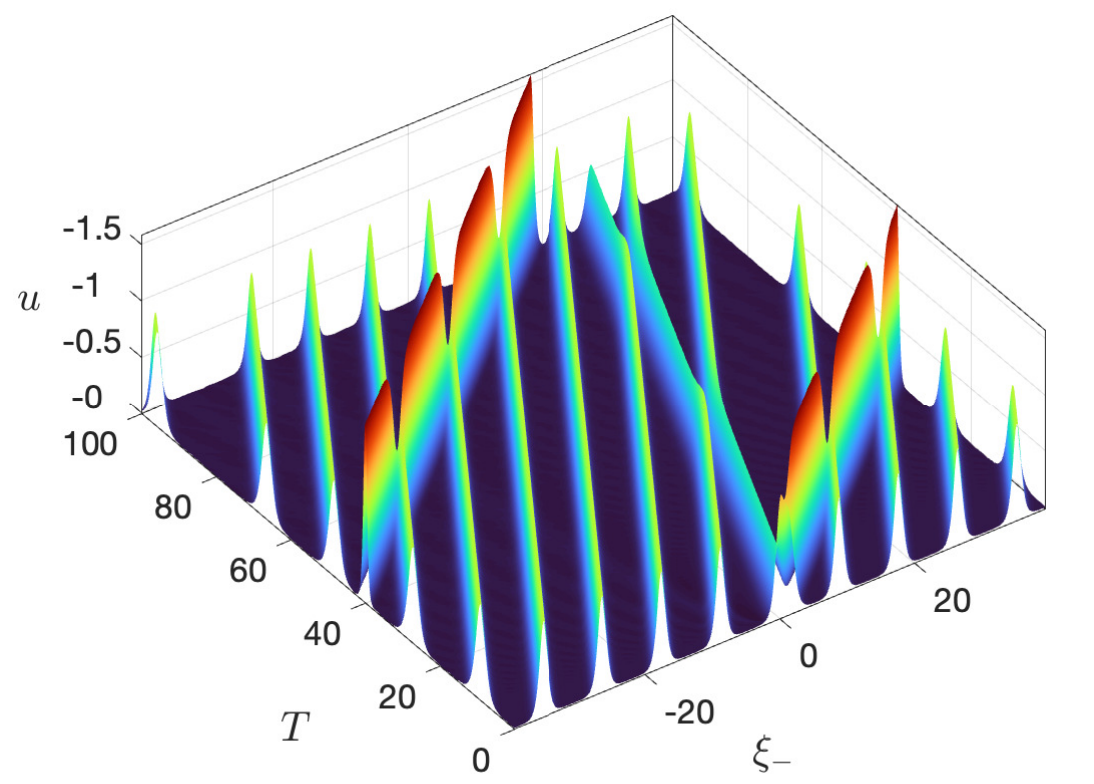}
    \vspace{0.25cm}
    \includegraphics[width=0.28\linewidth]{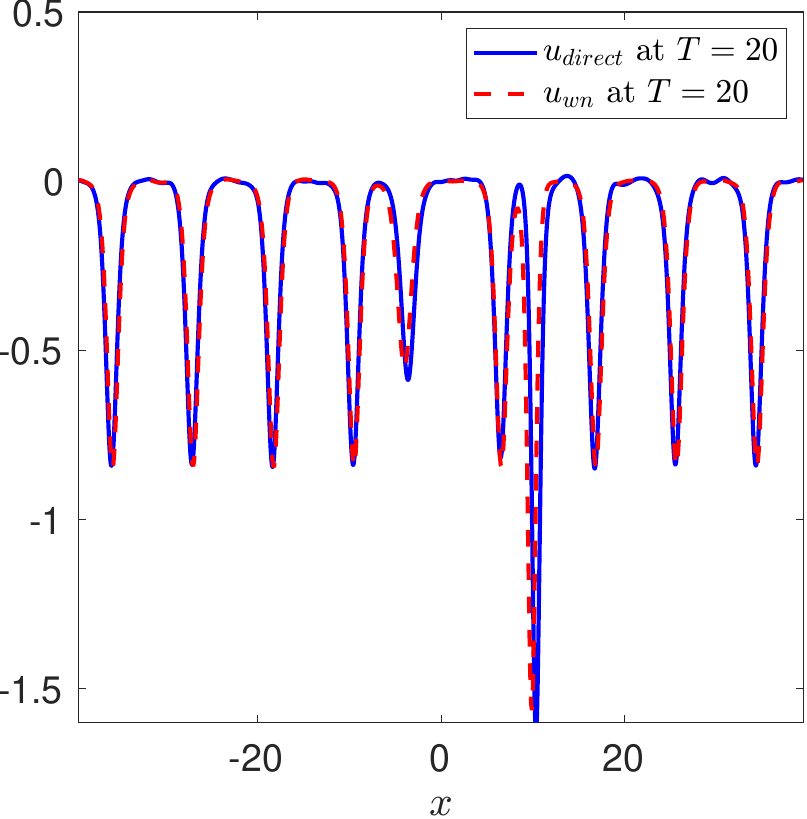} 
    \includegraphics[width=0.28\linewidth]{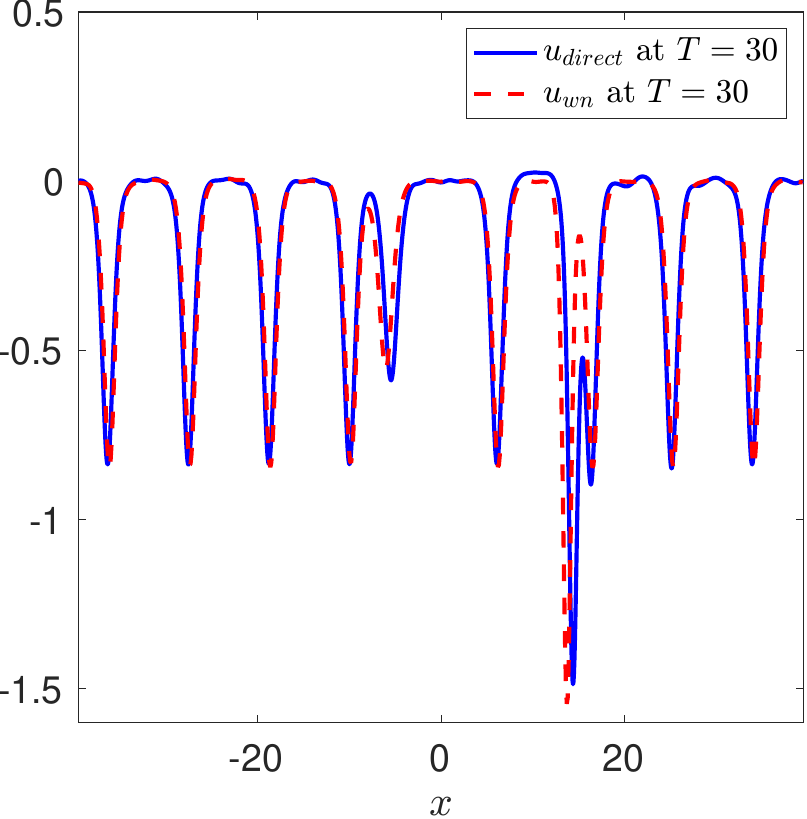}
    \includegraphics[width=0.28\linewidth]{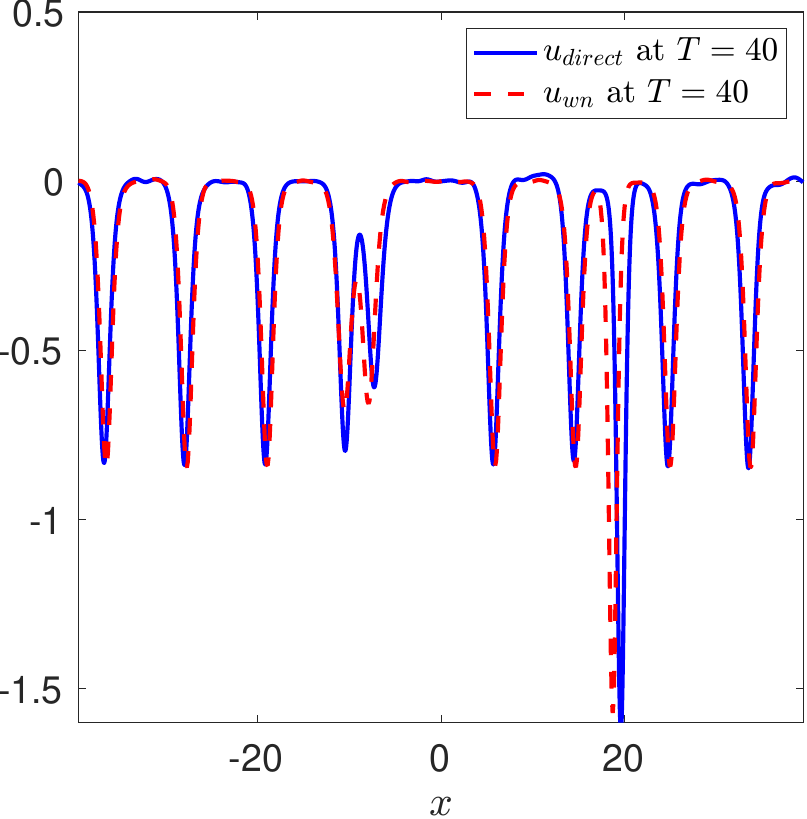}
     \vspace{0.25cm}
    \includegraphics[width=0.45\linewidth]{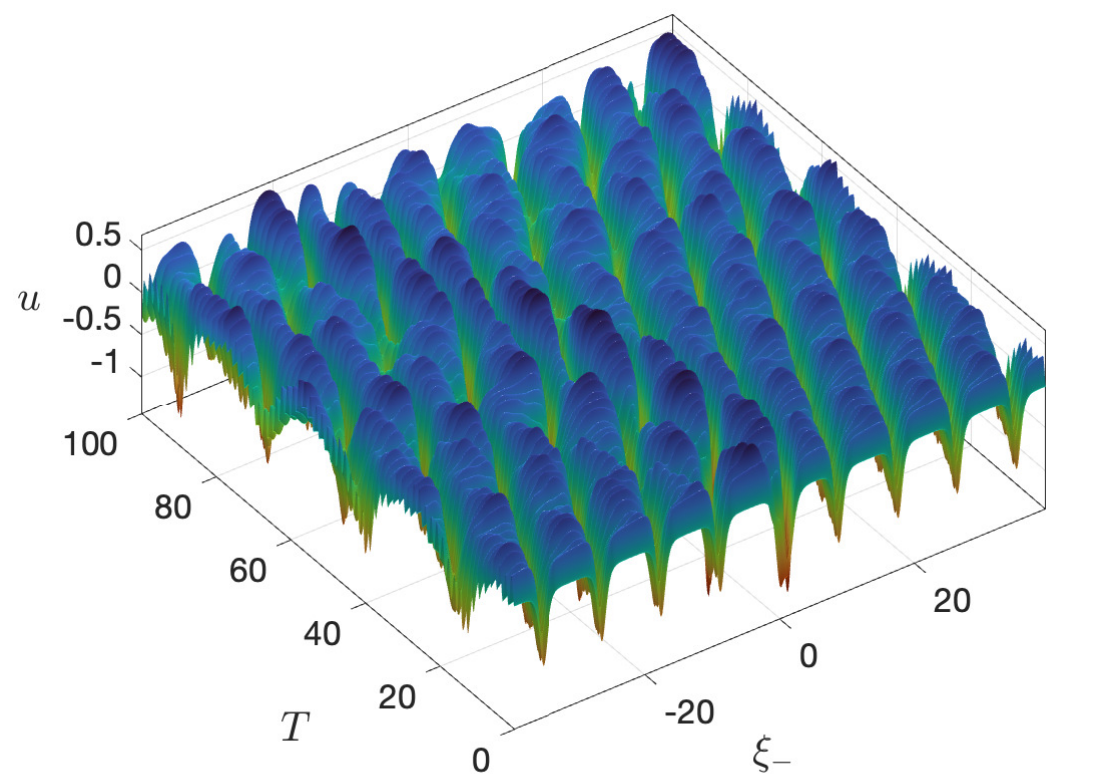}
    \includegraphics[width=0.45\linewidth]{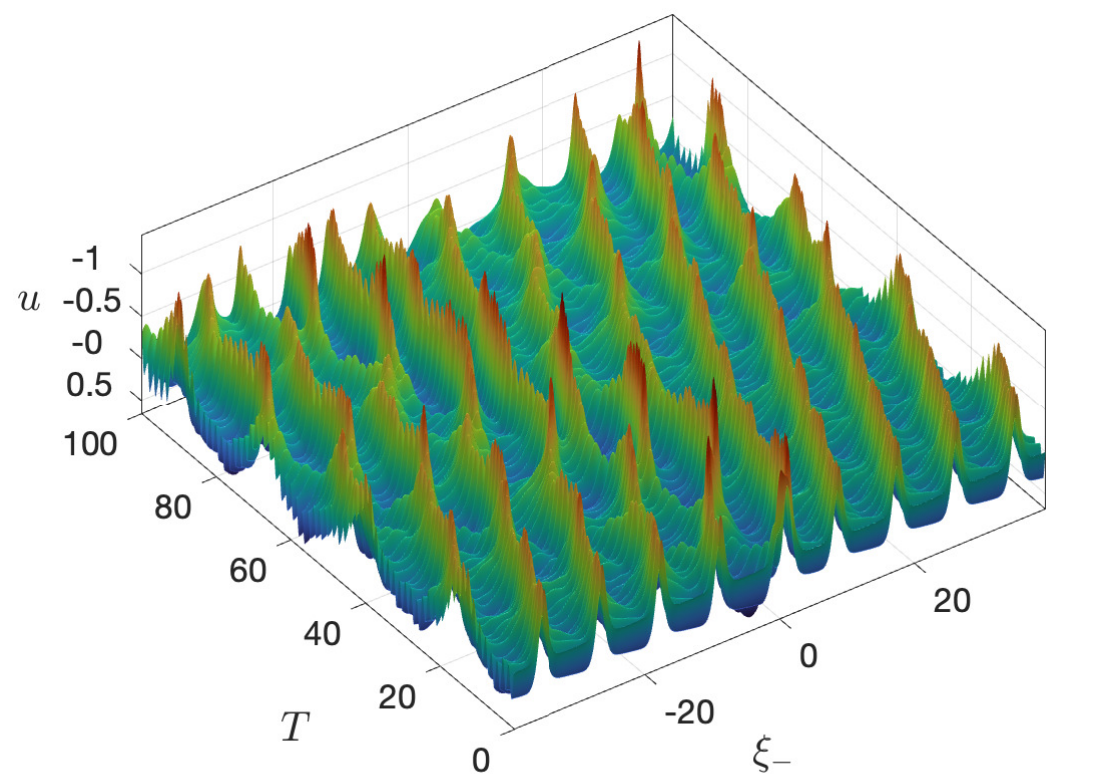}
    \vspace{0.25cm}
    \includegraphics[width=0.28\linewidth]{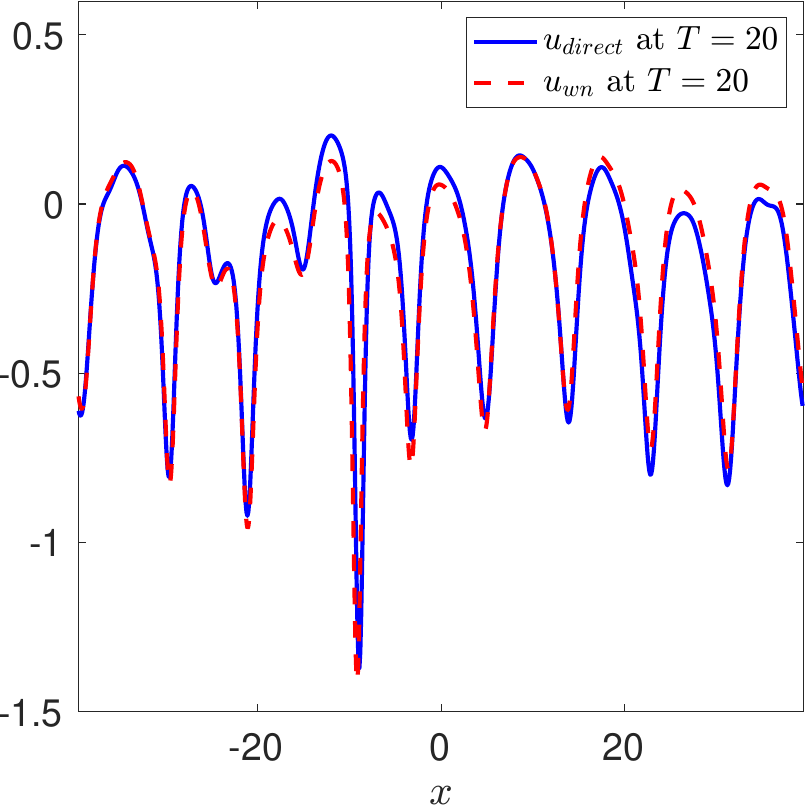} 
    \includegraphics[width=0.28\linewidth]{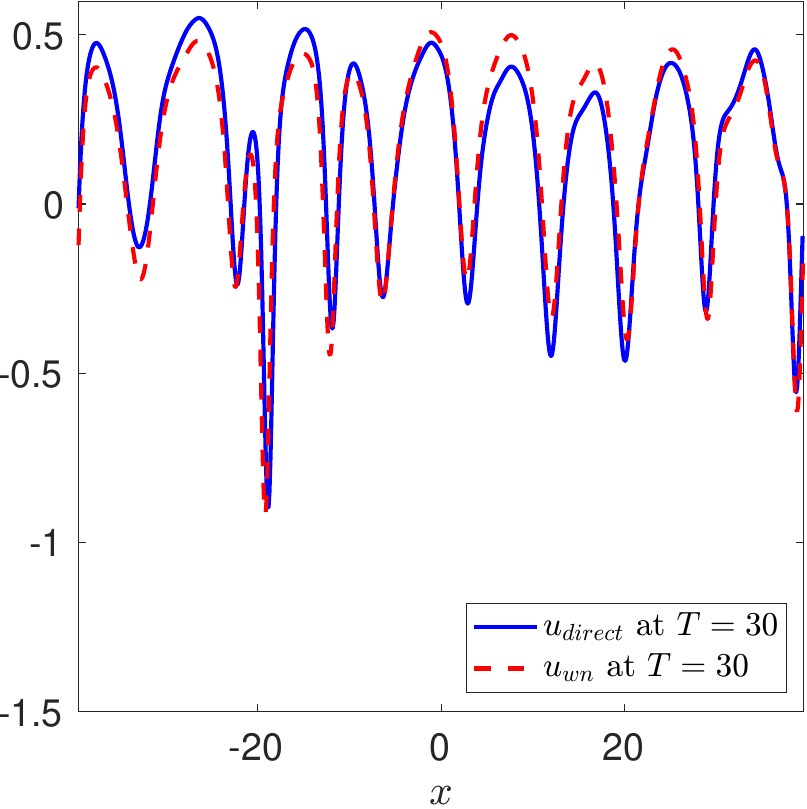}
    \includegraphics[width=0.28\linewidth]{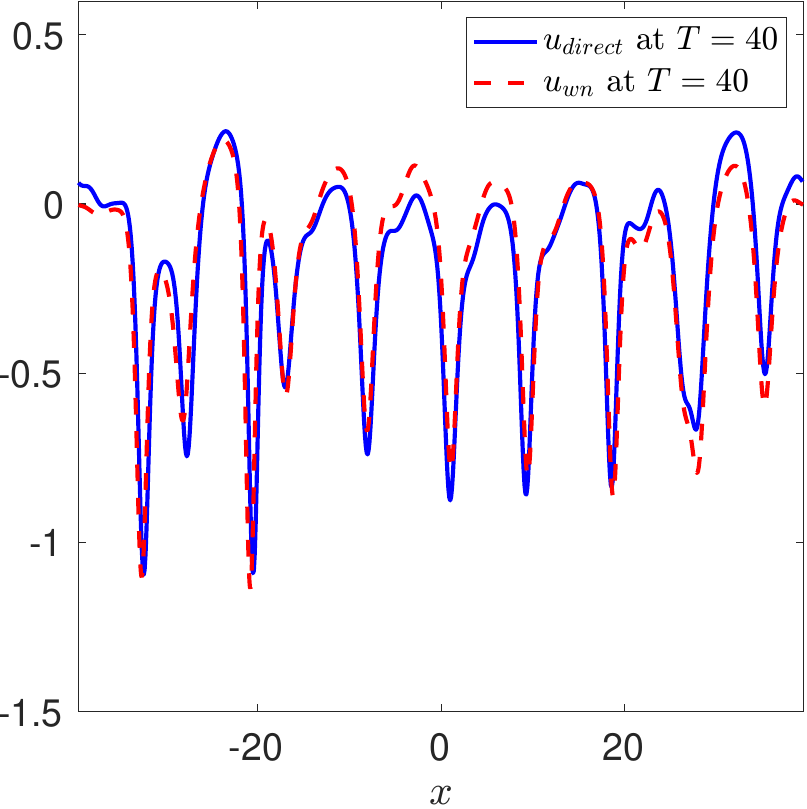}
    \caption{Evolution of the weakly-nonlinear solution \eqref{WNLFV} for a cnoidal wave initial condition with a generic localised perturbation, showing view from above (left) and below (right) for (a) $\gamma_1 = 0$, and (c) $\gamma_1 = 0.10$. Comparison of the direct numerical simulations (blue, solid) and weakly nonlinear solution (red, dashed) at times $T=20$ (left), $T=30$ (middle), and $T=40$ (right) for (b) $\gamma_1 = 0$, and (d) $\gamma_1 = 0.10$.  Numerical parameters are $\varepsilon = 0.005,\, \alpha_1 = -1.73,\, \beta_1 =  0.08, \, u_1 = -10^{-3},\, u_2=0,\, u_3=3$.} 
    \label{fig5.10}
\end{figure}

\begin{figure}
    \centering
    \includegraphics[width=0.43\linewidth]{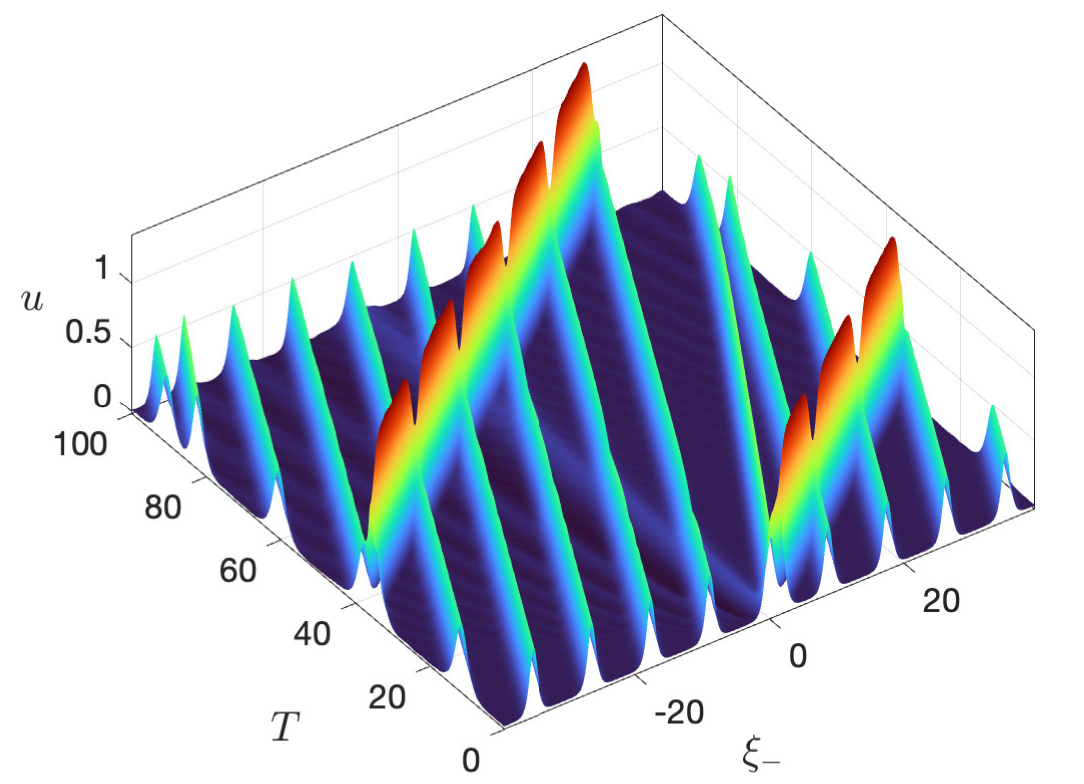}
    \includegraphics[width=0.43\linewidth]{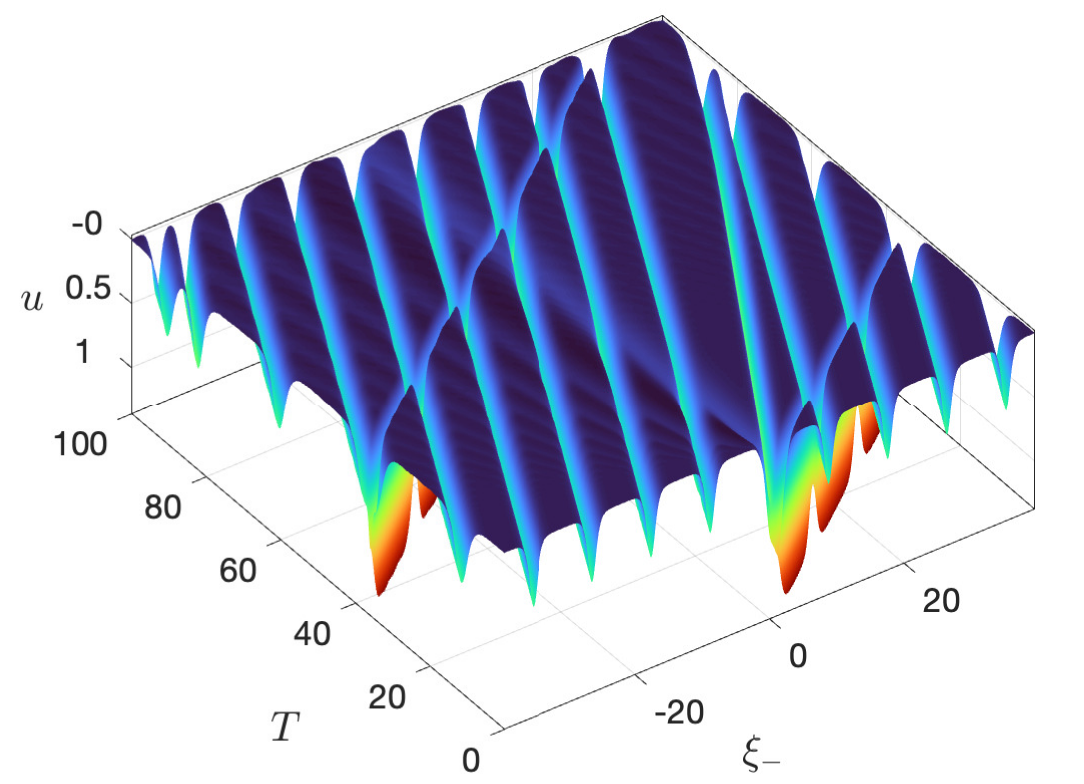}
    \vspace{0.25cm}
    \includegraphics[width=0.27\linewidth]{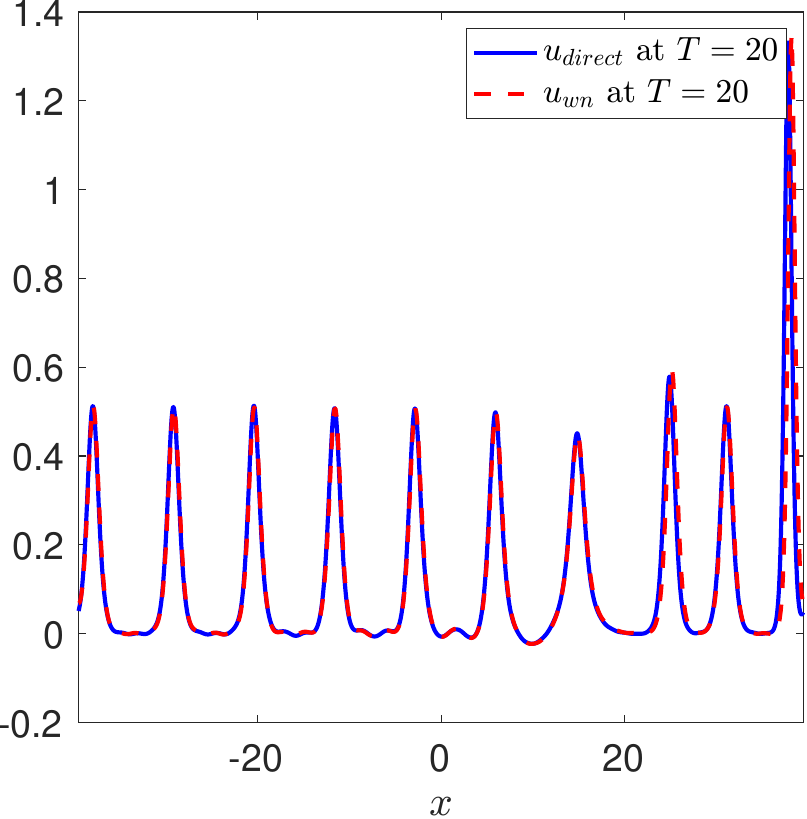} 
    \includegraphics[width=0.27\linewidth]{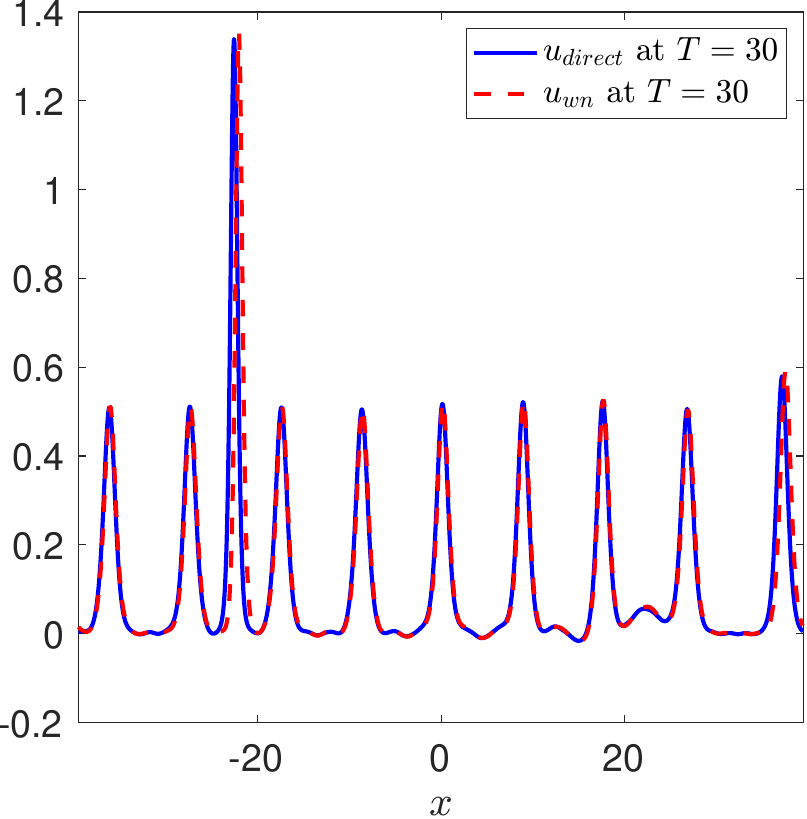}
    \includegraphics[width=0.27\linewidth]{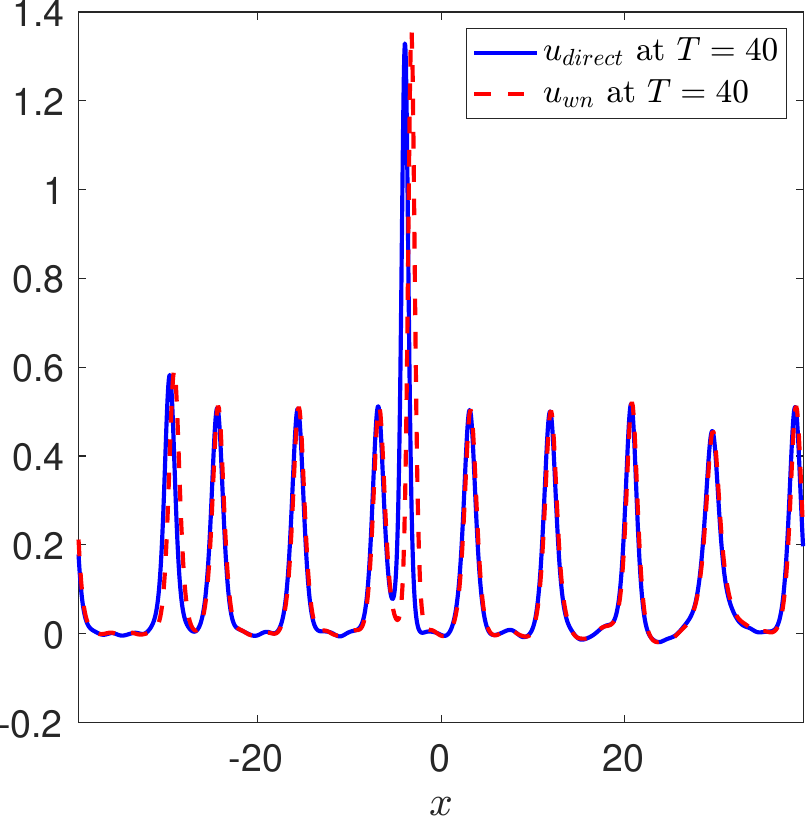}
    \vspace{0.25cm}
    \includegraphics[width=0.43\linewidth]{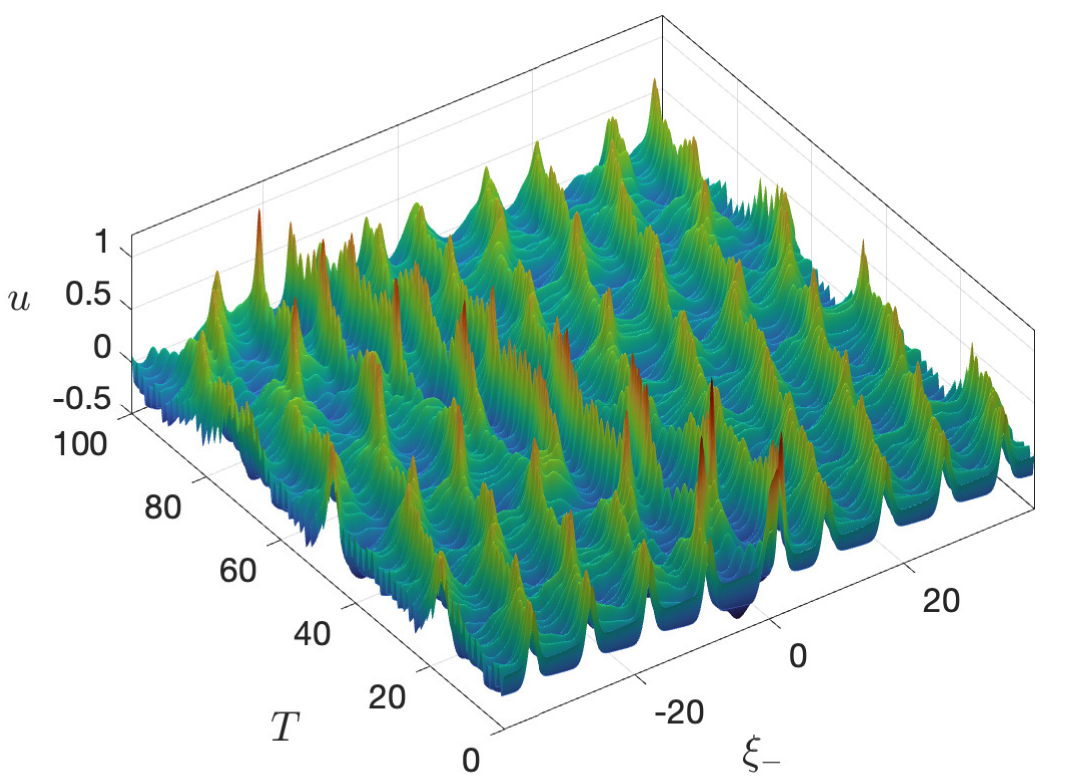}
    \includegraphics[width=0.43\linewidth]{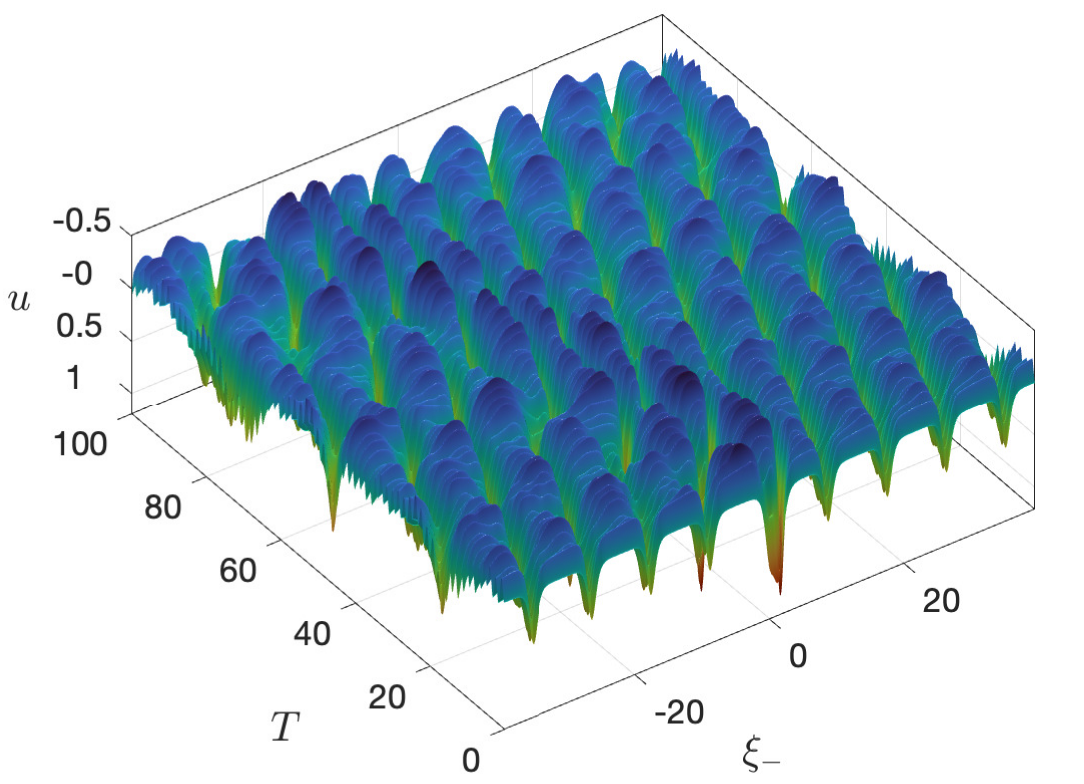}
    \vspace{0.25cm}
    \includegraphics[width=0.27\linewidth]{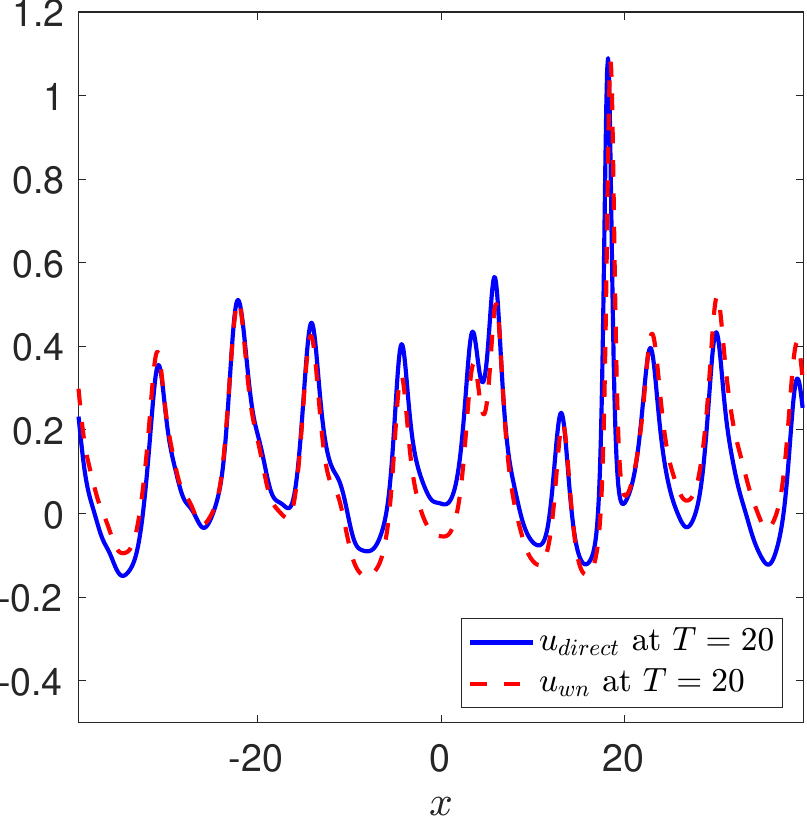} 
    \includegraphics[width=0.27\linewidth]{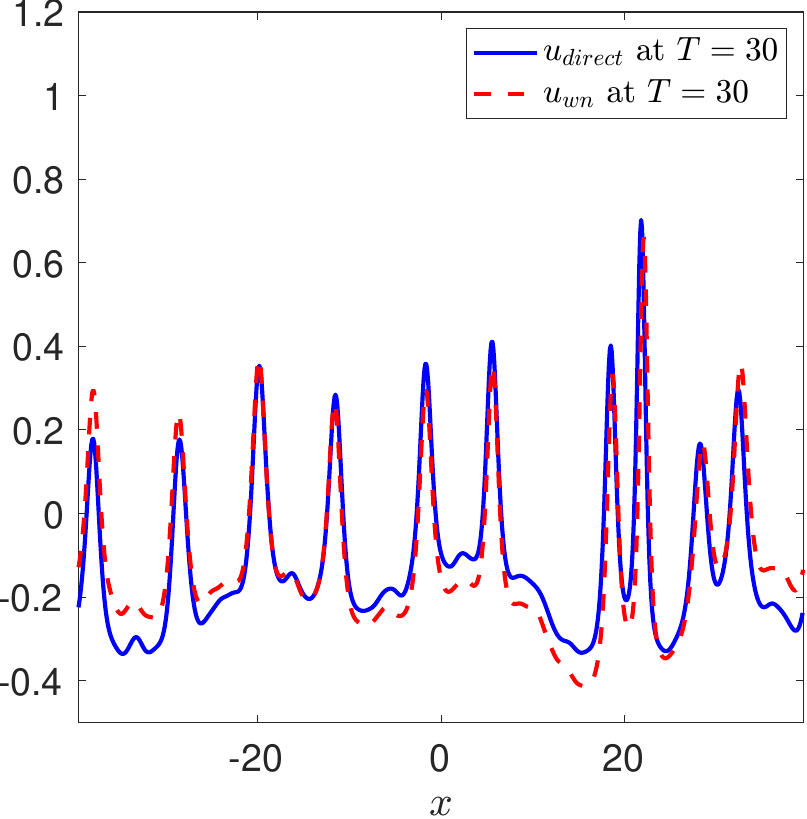}
    \includegraphics[width=0.27\linewidth]{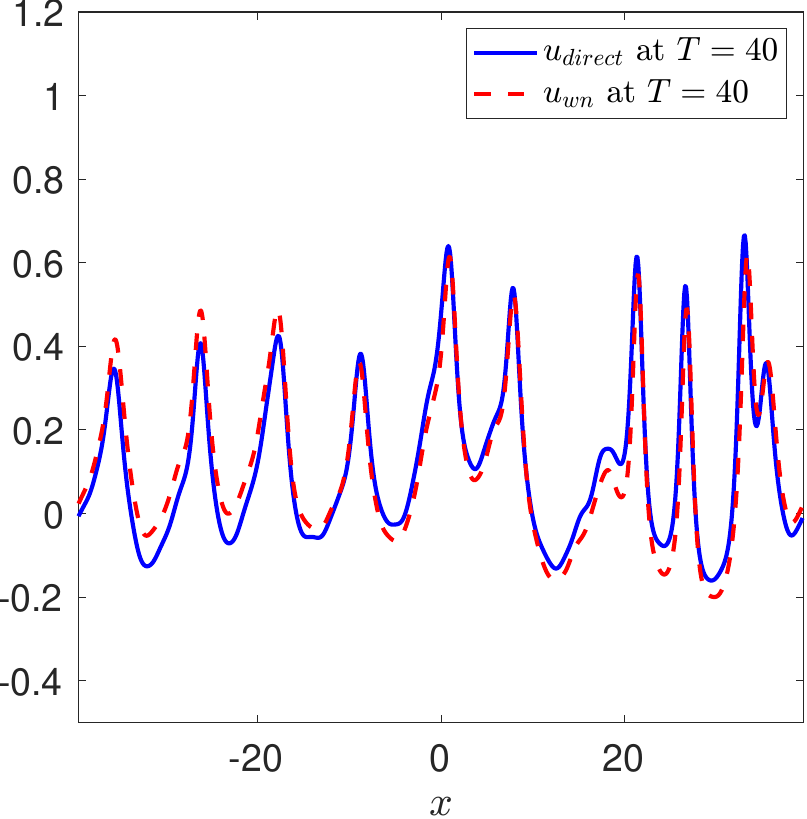}
    \caption{Evolution of the weakly-nonlinear solution \eqref{WNLFV} for a cnoidal wave initial condition with a generic localised perturbation, showing view from above (left) and below (right) for (a) $\gamma_1 = 0$, and (c) $\gamma_1 = 0.10$. Comparison of the direct numerical simulations (blue, solid) and weakly nonlinear solution (red, dashed) at times $T=20$ (left), $T=30$ (middle), and $T=40$ (right) for (b) $\gamma_1 = 0$, and (d) $\gamma_1 = 0.10$.  Numerical parameters are $\varepsilon = 0.005,\, \alpha_1 = 2.25,\, \beta_1 =  0.06, \, u_1 = -10^{-3},\, u_2=0,\, u_3=3$.
} 
    \label{fig5.13}
\end{figure}

\begin{figure}
    \centering
    \includegraphics[width=0.43\linewidth]{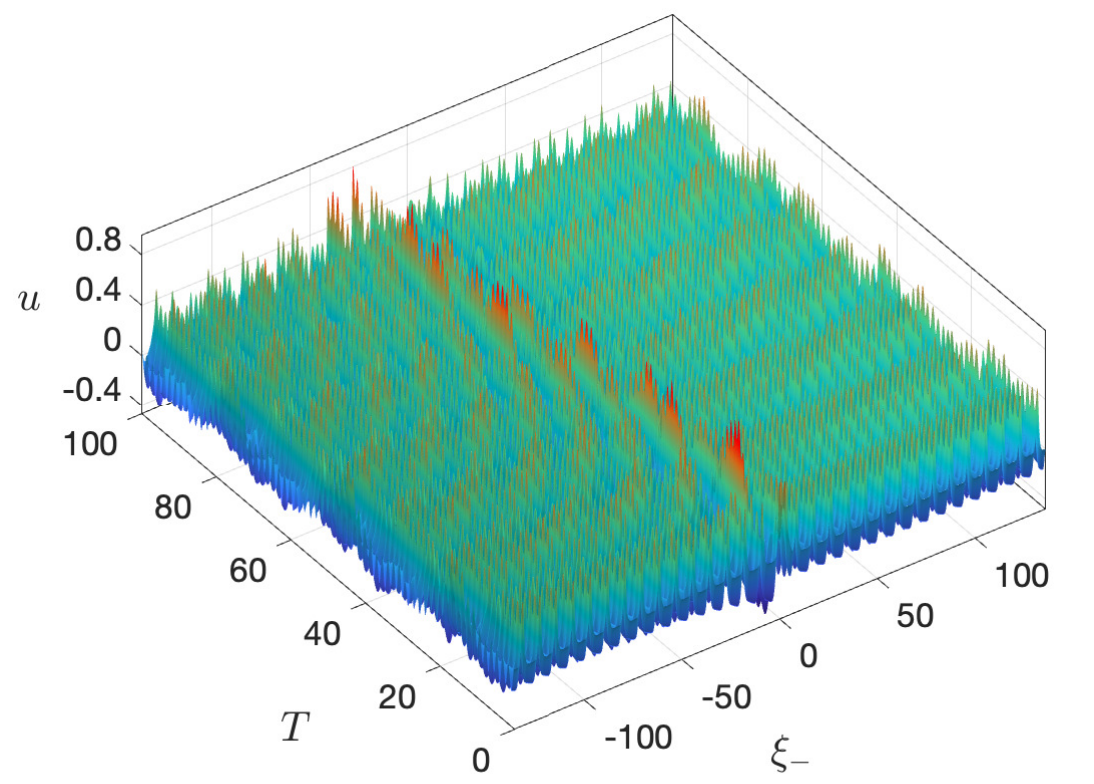} \ \
    \includegraphics[width=0.41\linewidth]{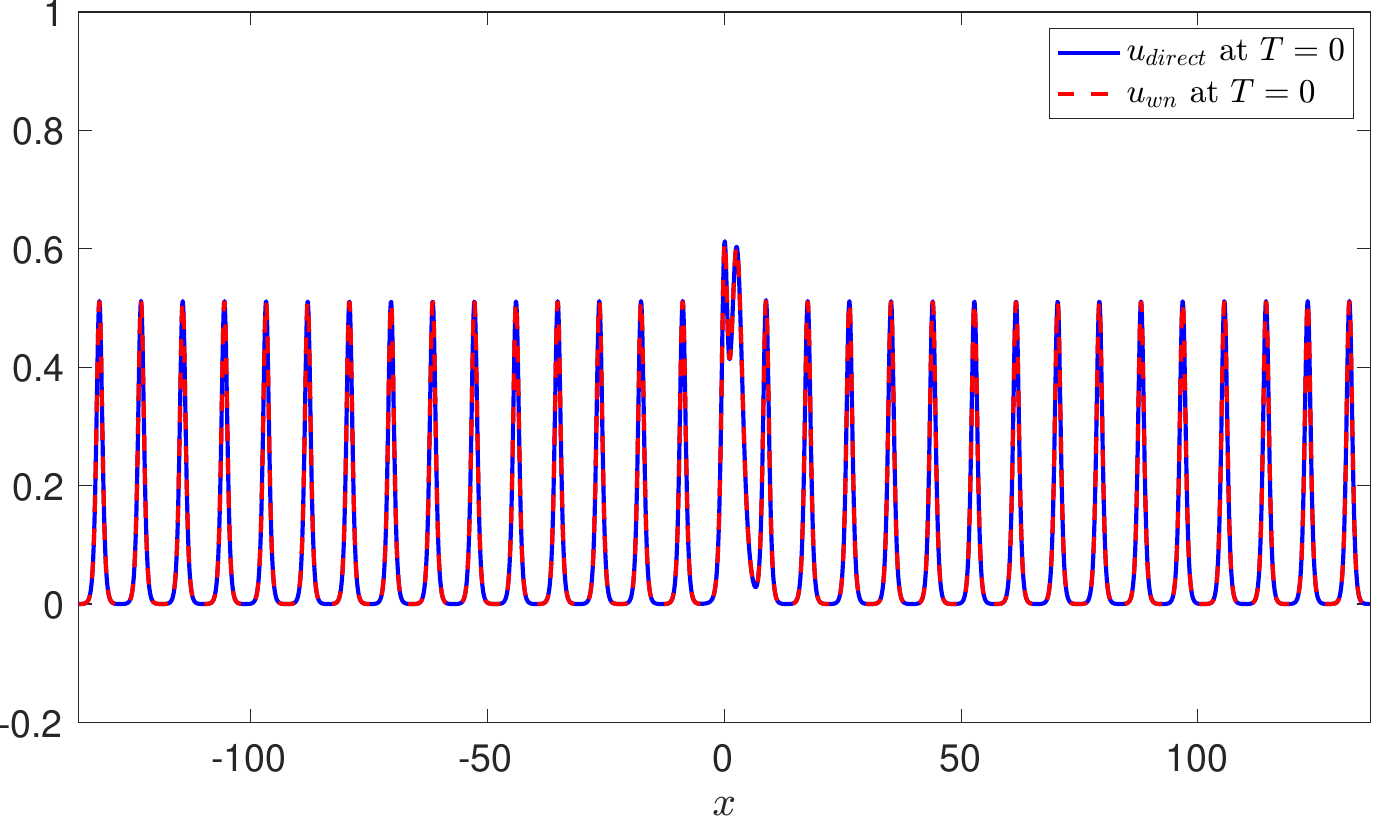}\\[1em]
    \includegraphics[width=0.41\linewidth]{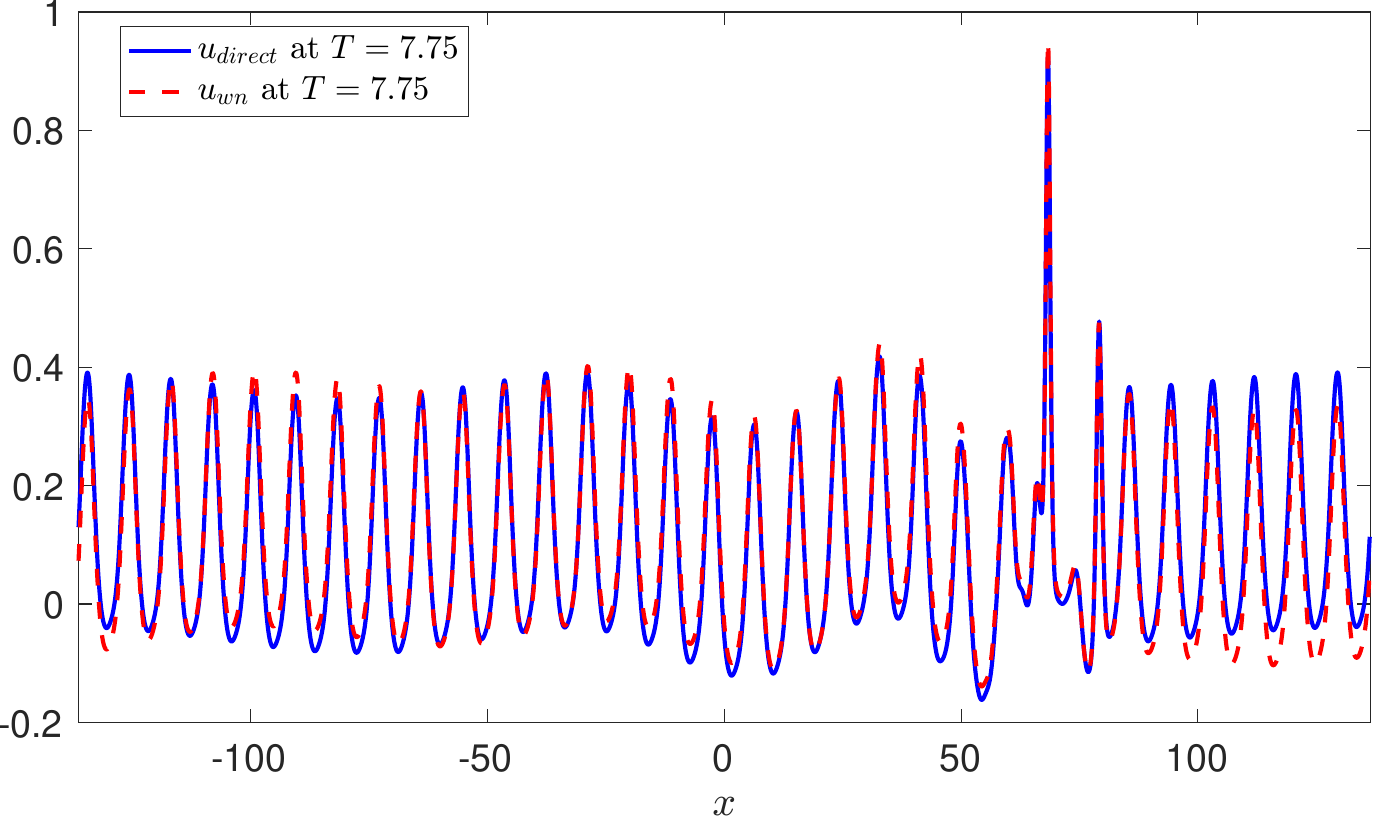}  \ \
    \includegraphics[width=0.41\linewidth]{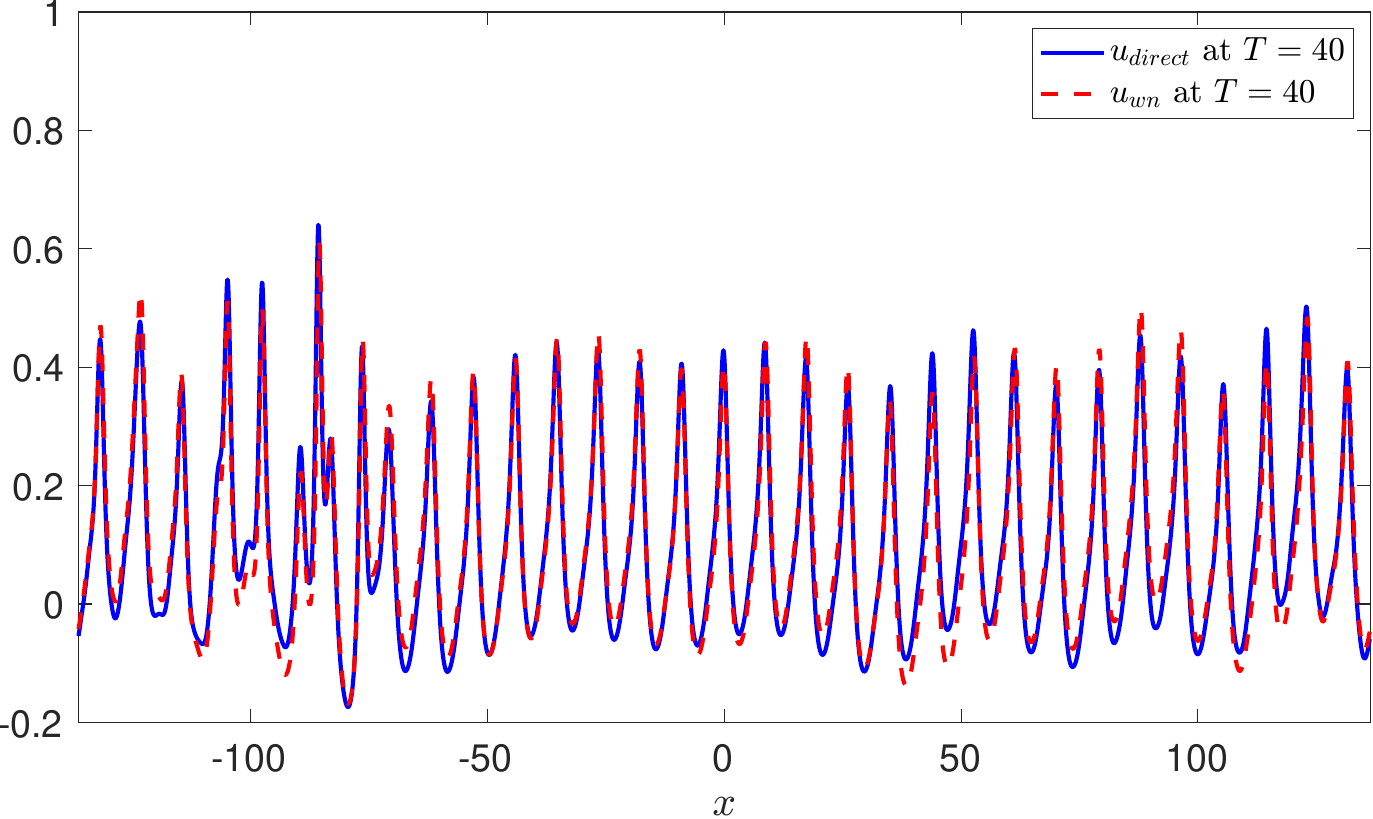}
    \caption{Rogue wave generation: (a) 3D view from above,  comparison of the direct numerical simulations (blue, solid) and weakly nonlinear solution (red, dashed) at times (b) $T=0$, (c) $T=7.75$, (d) $T=40$. Parameters are $\varepsilon = 0.005,\, \alpha_1 = 2.25,\, \beta_1 =  0.06,\, \gamma_1 = 0.10, \, u_1 = -10^{-3},\, u_2=0,\, u_3=3$.} 
    \label{fig5.17}
\end{figure}

	
\section{Concluding remarks}

In this study, we modelled the evolution of initial conditions in the form of cnoidal waves with periodicity defects and generic localised perturbations within the scope of the Boussinesq--Klein--Gordon (BKG) equation with periodic boundary conditions, extending our recent studies in \cite{NTK2025}. In \cite{NTK2025}, we considered Helfrich's (MMCC-$f$) extension \cite{H2007} of the Miyata--Maltseva--Choi--Camassa (MMCC) system \cite{Mi1988, Ma1989, CC1996, CC1999} for internal waves in a two-layer fluid with rotation. We constructed a weakly-nonlinear solution for uni-directional waves which led to the Ostrovsky equation, but avoided the zero-mean constraint on the initial conditions of the parent system, by deriving the Ostrovsky equation for deviations from generally non-zero and evolving mean values of the field variables. We used the constructed weakly-nonlinear solution to study the evolution of Cauchy problems with initial conditions of the same type as those discussed in our present study, with and without rotation. However, we did not test the constructed solutions against direct numerical simulations for the parent MMCC-$f$ system due to the difficulty of such simulations. Hence, in the present paper, our aim was to test the behaviours identified in \cite{NTK2025} within a simpler setting of the BKG equation.

The BKG equation is a physically relevant model that emerged in the studies of the long nonlinear bulk strain waves in solids, as discussed in the Introduction. The bi-directional weakly-nonlinear solution of this equation was constructed in \cite{KMP2014, KT2019}, using the Fourier series and d'Alembert's approaches, respectively.
We briefly overviewed the derivation developed in \cite{KT2019} in Section~2, to the accuracy including the first two leading orders in the square root of the small amplitude parameter. The Ostrovsky equations for the right- and left-propagating waves are derived for the deviations from the evolving mean values, and the zero-mean contradiction is avoided by the construction of the solution. We note that the construction of the weakly-nonlinear solution for internal waves in \cite{NTK2025} had a significant complication compared to this derivation, since the equations for the mean fields were coupled to the equations for the deviations. However, we showed that it is possible to develop simultaneous asymptotic expansions for the mean-field variables and the deviations (see \cite{NTK2025} for the details).

Hence, in the present study we compared the constructed weakly-nonlinear solution with direct numerical simulations for the BKG equation both for $\gamma_1 = 0$ (KdV regime) and $\gamma_1 \ne 0$ (Ostrovsky regime) for the initial conditions corresponding to the pure cnoidal waves of the KdV equation (Section 3), cnoidal waves with expansion defects (Section 4) and cnoidal waves with generic localised perturbations (Section~5). In all cases, there was very good agreement between the weakly-nonlinear solutions and the direct numerical simulations. Moreover, the direct numerical simulations inevitably contained small counter-propagating perturbations, and our modelling has shown stability of all uni-directional solutions and evolution scenarios discussed in this paper. 

The main conclusions can be summarised as follows. In the KdV regime ($\gamma_1 = 0$), cnoidal waves with localised perturbations generally lead to the emergence of bright and dark breathers on a cnoidal-wave background, and expansion and contraction defects, both in the weakly-nonlinear solution, and in the full bi-directional BKG equation. The cnoidal waves with expansion or contraction defects  constitute generalised (shock-like) travelling waves of the KdV equation, satisfying all (infinitely many) conservation laws of the KdV equation and a natural weak formulation; moreover, the cnoidal wave with an expansion defect is a broken extremal of the associated variational problem \cite{NTK2025}.  Both expansion and contraction defects were long-lived in our simulations. In the Ostrovsky regime ($\gamma_1 \ne 0$), all these wave components lead to the emergence of strong bursts of energy slowly propagating to the left (in the reference frame moving with the linear long-wave speed). Hence, the qualitative outcome is similar to that reported for the initial conditions in the form of solitons and slowly-modulated cnoidal waves in the earlier studies (see, for example, \cite{GH2008, WJ2014, WJ2017, S2020} and references therein), but the formation of the bursts happens much faster (in the comparisons discussed in \cite{NTK2025}, it was tens of times faster).

Finally, we would like to note a numerical experiment, based on the constructed weakly-nonlinear solution, where the initial condition was extended from the $[-L, L]$ interval to a bigger domain, where it was gradually decreasing to zero near the boundaries ($\kappa= 0.5$):
\begin{eqnarray}
   && u|_{t=0} = \frac{1}{2} \left[ \tanh \left( \kappa (x + L/2) \right) -\tanh \left( \kappa (x - L/2) \right)\right] \times \nonumber \\
    && \left \{ \dfrac{6\beta_1}{\alpha_1}\Bigg ( u_2 + (u_3 - u_2) \text{ cn}^2\Big[ \xi_-  \sqrt{\dfrac{u_3 -u_1}{2}} ; m \Big] \Bigg )_{t=0}  + A_1 \text{sech}^2 \Big[ A_2(x -\xi_0) \Big] \right \}.
    \label{cnoidal_ad}
\end{eqnarray}
The parameters used in these runs were $\varepsilon = 0.005, \alpha_1 = - 1.73, \beta_1 = 0.08, \gamma_1 = 0.10, u_1 = - 10^{-3}, u_2 = 0, u_3 = 3, A_1 = -0.8, A_2 = 1, \xi_0 = 1.3$.   Hence, this modelled the case of data on the infinite line with solutions decaying at infinity, and numerical simulations were performed with zero boundary conditions and sponge layers near the boundaries (see, for example, \cite{AGK2013} for the discussion of the sponge layers). The results are shown in the first two rows of Figure \ref{R}, and compared with the solution from our manuscript, computed using the initial condition
\begin{eqnarray}
    u|_{t=0} = 
     \dfrac{6\beta_1}{\alpha_1}\Bigg ( u_2 + (u_3 - u_2) \text{ cn}^2\Big[ \xi_-  \sqrt{\dfrac{u_3 -u_1}{2}} ; m \Big] \Bigg )_{t=0}
    + A_1 \text{sech}^2 \Big[ A_2(x -\xi_0) \Big],
    \label{cnoidal_ad1}
\end{eqnarray}
with periodic boundary conditions. These numerical simulations show that there exists a considerable period of time such that the behaviour in the central part of the large domain can be rather accurately approximated by the simulations on a much smaller domain, with periodic boundary conditions, which are shown in the third row of Figure \ref{R}. The evolution scenarios discussed in the paper are relevant to the behaviour in this part.

\begin{figure}[h]
	\centering
	\includegraphics[width=0.45\linewidth]{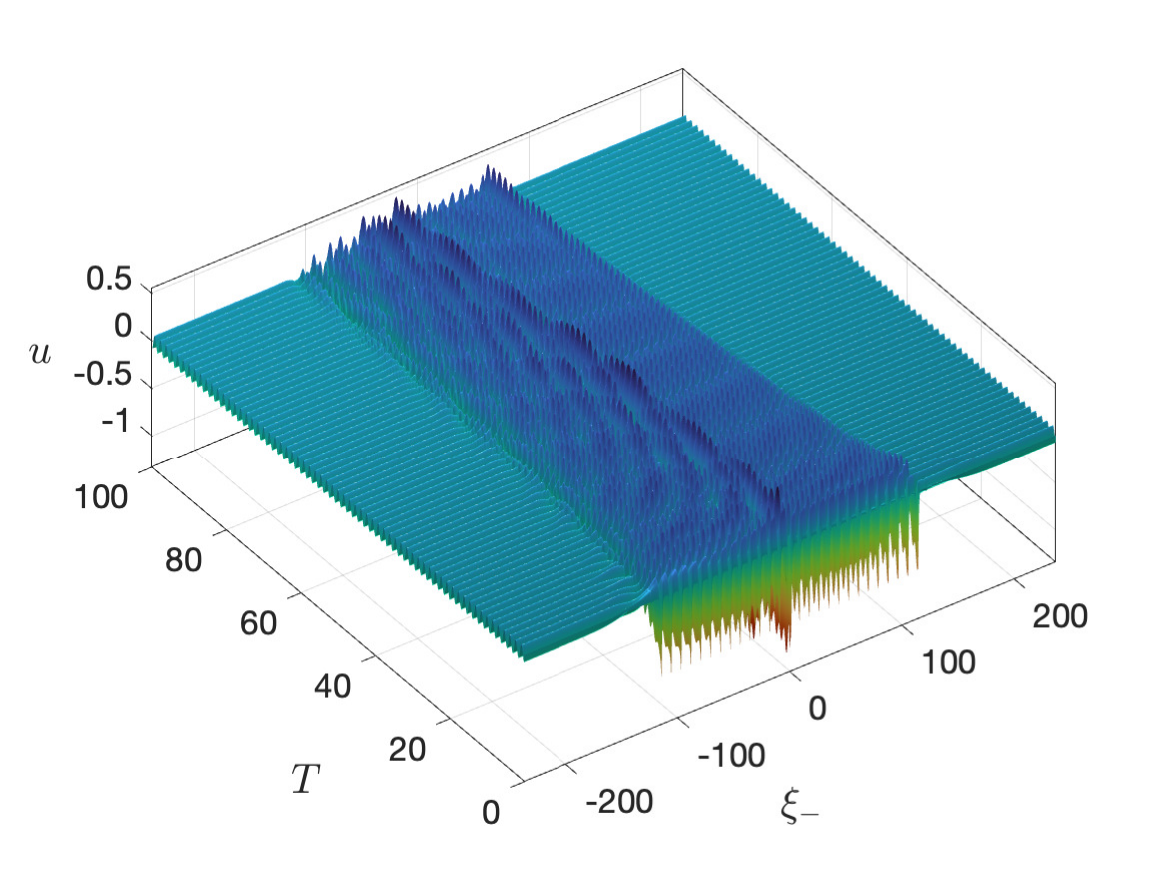} 
   	\includegraphics[width=0.45\linewidth]{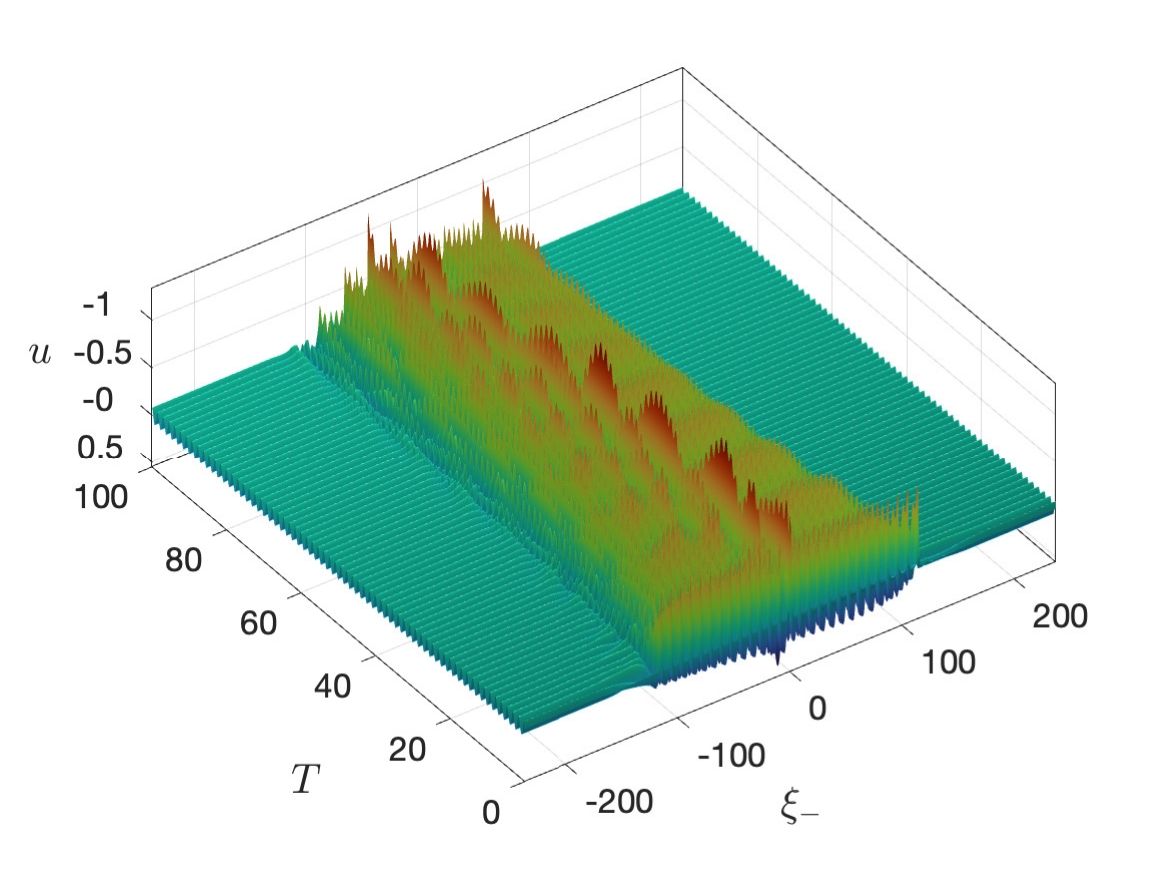} \\
	\includegraphics[width=0.45\linewidth]{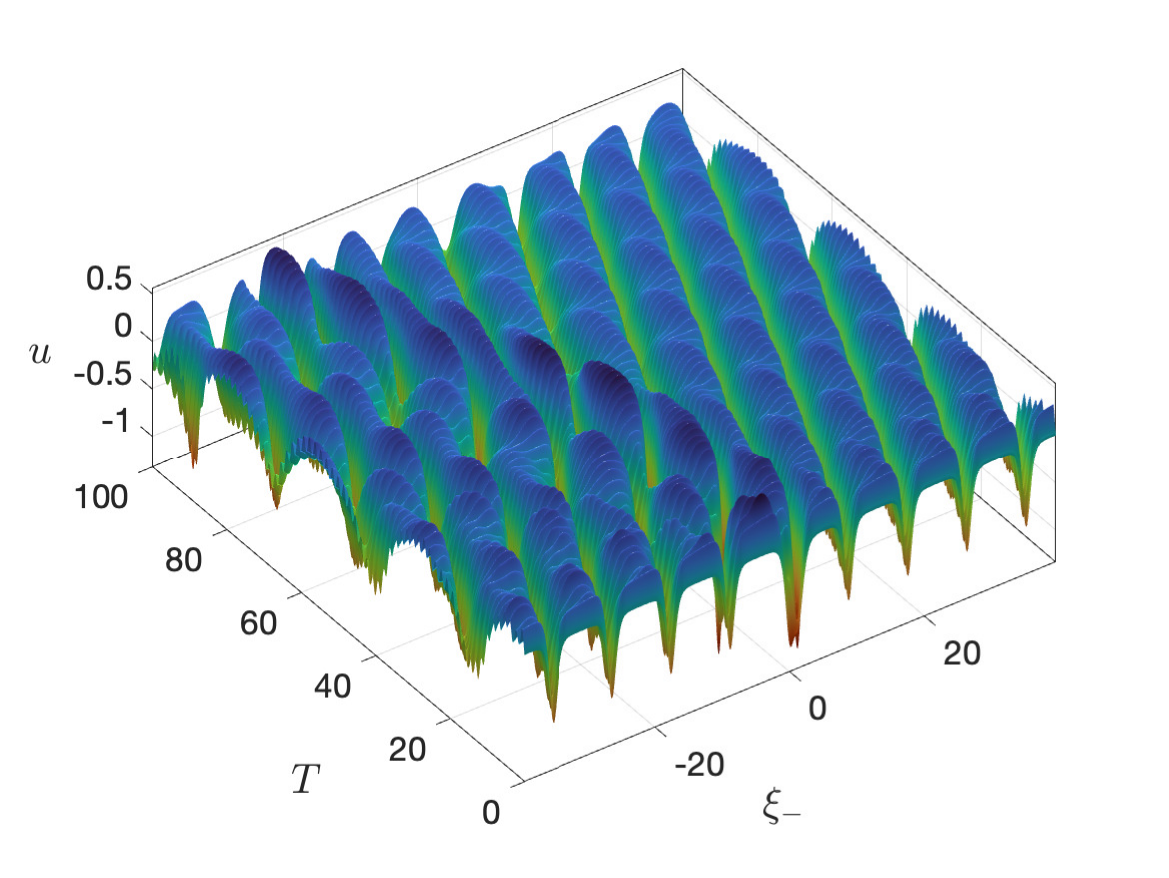} 
   	\includegraphics[width=0.45\linewidth]{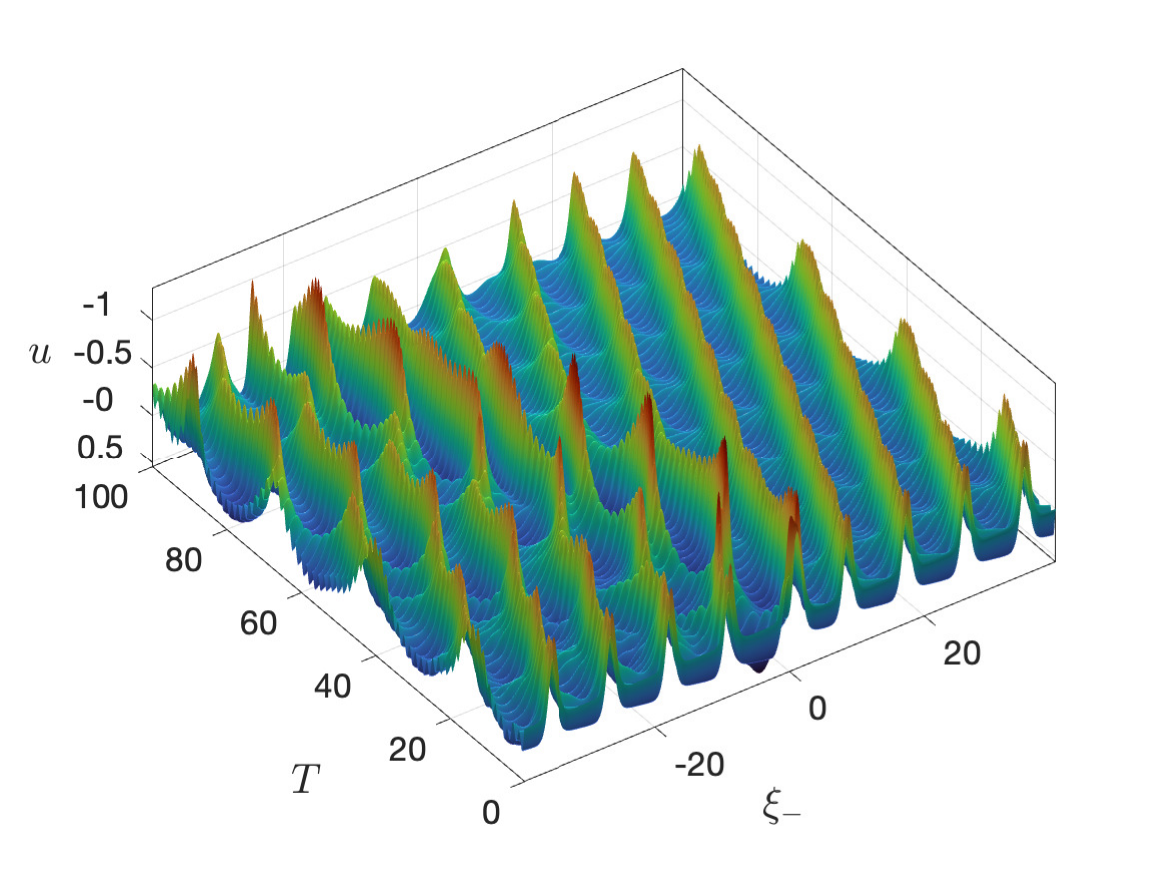} \\
	\includegraphics[width=0.43\linewidth]{Figures/xi_u_gen_lo_rot_on_3D-crop.pdf} 
	\includegraphics[width=0.43\linewidth]{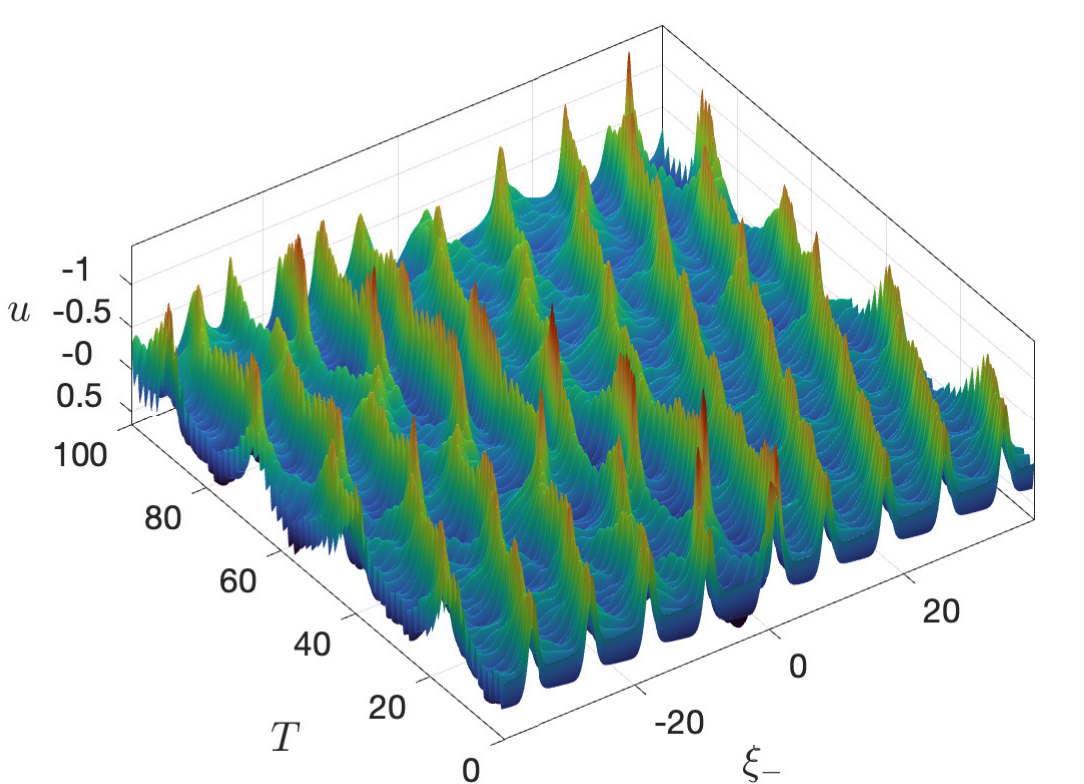} 
 
	\caption{First and second rows: Evolution of the initial condition (\ref{cnoidal_ad}) in computations with zero boundary conditions on a large domain: view from above (left) and below (right). Third row: Evolution of the initial condition (\ref{cnoidal_ad1})  in simulations with periodic boundary conditions on a smaller domain: view from above (left) and below (right). Numerical parameters are $\varepsilon = 0.005,\, \alpha_1 = -1.73,\, \beta_1 =  0.08, \gamma_1 = 0.10, \, u_1 = -10^{-3},\, u_2=0,\, u_3=3$.	}
	\label{R}
	\end{figure}


\clearpage
\begin{thebibliography}{00}
\bibitem{O1978}
L.A. Ostrovsky, Nonlinear internal waves in a rotating ocean, \textit{Oceanology} 18 (1978) 119-125.
\bibitem{G1985}
R.H.J. Grimshaw, Evolution equations for weakly nonlinear, long internal waves in a rotating fluid, \textit{Stud. Appl. Maths}  73 (1985) 1-33.
\bibitem{OS1998}
M.A. Obregon, Yu.A. Stepanyants, Oblique magneto-acoustic solitons in rotating plasma, \textit{Phys. Lett. A} 249 (1998) 315-323.
\bibitem{KSZ2009}
K.R. Khusnutdinova, A.M. Samsonov, A.S. Zakharov, Nonlinear layered lattice model and generalized solitary waves in imperfectly bonded structures, \textit{Phys. Rev. E} 79 (2009) 056606.
\bibitem{KM2011}
K.R. Khusnutdinova, K.R. Moore, Initial-value problem for coupled Boussinesq equations and a hierarchy of Ostrovsky equations, \textit{Wave Motion} 48 (2011) 738-752.
\bibitem{KT2017}
K.R. Khusnutdinova, M.R. Tranter, On radiating solitary waves in bi-layers with delamination and coupled Ostrovsky equations, \textit{Chaos} 27 (2017) 013112.
\bibitem{KT2022}
K.R. Khusnutdinova, M.R. Tranter, Periodic solutions of coupled Boussinesq equations and
Ostrovsky-type models free from zero-mass contradiction, \textit{Chaos} 32 (2022) 113132.
\bibitem{TCPT2024}
J.S. Tamber, D.J. Chappell, J.C. Poore, M.R. Tranter, Detecting delamination via nonlinear wave scattering in a bonded elastic bar, \textit{Nonlinear Dyn.} 112 (2024) 23-33.
\bibitem{TCT2025}
J.S. Tamber, D.J. Chappell, M.R. Tranter, Delamination detection in layered waveguides using
Ostrovsky wave packets, \textit{Proc. Roy. Soc. A} 481 (2025) 20240574.
\bibitem{AGK2013}
A. Alias, R.H.J. Grimshaw, K.R. Khusnutdinova, On strongly interacting internal waves in a rotating ocean and coupled Ostrovsky equations, \textit{Chaos} 23 (2013) 023121.
\bibitem{AGK2014}
A. Alias, R.H.J. Grimshaw, K.R. Khusnutdinova, Coupled Ostrovsky equations for internal waves in a shear flow, \textit{Phys. Fluids}  26 (2014) 126603.
\bibitem{GPTK2010}
R. Grimshaw, E.N. Pelinovsky, T.G. Talipova, O.E. Kurkina, Internal solitary waves: propagation, deformation and disintegration, \textit{Nonlin. Processes Geophys.} 17 (2010) 633-649.
\bibitem{NSS2015}
S. Nikitenkova, N. Singh, Y. Stepanyants, Modulational stability of weakly nonlinear wave-trains in media with small- and large-scale dispersions, \textit{Chaos} 25 (2015) 123113.
\bibitem{GSM2015}
R. Grimshaw, J.C.B. da Silva, J.M. Magalhaes, Modelling and observations of oceanic nonlinear internal wave packets affected by the Earth's rotation, \textit{Ocean Modelling} 116 (2017) 146-158.
\bibitem{ORS2018}
M. Obregon, N. Raj, Y. Stepanyants, Adiabatic decay of internal solitons due to Earth's rotation within the framework of the Gardner-Ostrovsky equation, \textit{Chaos} 28 (2018) 033106.
\bibitem{G2015}
R. Grimhsaw, Effect of a background shear current on models for nonlinear long internal waves, \textit{Fundam. Appl. Hydrophysics} 8  (2015) 20-23.
\bibitem{KMP2014}
K.R. Khusnutdinova, K.R. Moore, D.E. Pelinovsky, Validity of the weakly nonlinear solution of the Cauchy problem for the Boussinesq-type equation, \textit{Stud. Appl. Maths} 133 (2014) 52-83.
\bibitem{KT2019}
K.R. Khusnutdinova, M.R. Tranter, D’Alembert-type solution of the Cauchy problem for the
Boussinesq-Klein-Gordon equation, \textit{Stud. Appl. Maths} 142 (2019) 551-585.
\bibitem{NTK2025}
K. Nirunwiroj, D. Tseluiko, K. Khusnutdinova, Evolution of internal cnoidal waves with local defects with and without rotation, \textit{J. Fluid Mech.} 1023 (2025) A26.
\bibitem{H2007}
K.R. Helfrich, Decay and return of internal solitary waves with rotation. \textit{Phys. Fluids} 19 (2007) 026601.
\bibitem{Mi1988}
M. Miyata, Long Internal Waves of Large Amplitude, In \textit{ Nonlinear Water Waves}, Springer (1988)  399-406.
\bibitem{Ma1989}
Z.L. Maltseva, Unsteady long waves in a two-layer fluid. In \textit{Dinamika Sploshn. Sredy}, vol. 93 (1989)
96–110 (in Russian).
\bibitem{CC1996}
W. Choi, R. Camassa,  Weakly nonlinear internal waves in a two-fluid system, \textit{J. Fluid Mech.} 313 (1996) 83-103.
\bibitem{CC1999}
W. Choi, R. Camassa, Fully nonlinear internal waves in a two-fluid system, \textit{J. Fluid Mech.}  396 (1999)
1-36.
\bibitem{GOSS1998}
R.H.J. Grimshaw, L.A. Ostrovsky, V.I. Shrira, Y.A. Stepanyants, Long nonlinear surface and internal gravity waves in a rotating ocean, \textit{Surv. Geophys.} 19 (1998) 289-338.
\bibitem{GH2012}
R. Grimshaw, K. Helfrich, The effect of rotation on internal solitary waves, \textit{IMA J. Appl. Math.} 77 (2012) 326-339.
\bibitem{S2020}
Y.A. Stepanyants, Nonlinear waves in a rotating ocean (The Ostrovsky equation and its generalizations
and applications), \textit{Izv. Atmos. Ocean. Phys} 56 (2020) 16-32.
\bibitem{L1981}
A.I. Leonov, The effect of the Earth's rotation on the propagation of weak nonlinear surface and internal long oceanic waves, \textit{Ann. NY Acad. Sci.} 373 (1981) 150-159.
\bibitem{GS1991}
V.N. Galkin, Yu.A. Stepanyants, On the existence of stationary solitary waves in a rotating fluid, \textit{J. Appl. Maths. Mech.} 55 (1991) 939-943.
\bibitem{GHO1998}
R.H.J. Grimshaw, J.-M. He, L. A. Ostrovsky, Terminal damping of a solitary wave due to radiation in rotational systems, \textit{Stud. Appl. Maths} 101 (1998) 197-210.
\bibitem{GH2008}
R. Grimshaw, K. Helfrich, Long-time solutions of the Ostrovsky equation, \textit{Stud. Appl. Maths} 121 (2008) 71-88.
\bibitem{GHJ2013}
R.H.J. Grimshaw, K.R. Helfrich, E.R. Johnson, Experimental study of the effect of rotation on large amplitude internal waves,  \textit{Phys. Fluids} 25 (2013) 056602.
\bibitem{YK2001}
D. Yagi, T. Kawahara, Strongly nonlinear envelope soliton in a lattice model for periodic structure, \textit{Wave Motion} 34 (2001) 97-107.
\bibitem{WJ2014}
A.J. Whitfield, E.R. Johnson, Rotation-induced nonlinear wavepackets in internal waves, \textit{Phys. Fluids} 26 (2014) 056606.
\bibitem{WJ2017}
A.J. Whitfield, E.R. Johnson, Whitham modulation theory for the Ostrovsky equation. \textit{Proc. R. Soc. A} 473 (2017) 20160709.
\bibitem{JOP2025}
M.A. Johnson, J. Oregano, W.R. Perkins, On the modulation of wave trains in the Ostrovsky
equation, arXiv:2505.21466 [math.AP] (2025) 42pp.
\bibitem{GNST2020}
S. Gavrilyuk, B. Nkonga, K.-M. Shyue, L. Truskinovsky,  Stationary shock-like transition fronts
in dispersive systems, {\it Nonlinearity} 33 (2020) 5477-5509.
\bibitem{GS2022}
S. Gavrilyuk,K.-M. Shyue,  Singular solutions of the BBM equation: analytical. and numerical
study, \textit{Nonlinearity} 35 (2022) 388-410.
\bibitem{GNS2024}
S. Gavrilyuk, B. Nkonga, K.-M. Shyue,  The conduit equation: hyperbolic approximation and generalized Riemann problem. {\it J. Comp. Phys.} 2024 (2024) 113232.
\bibitem{KM1975}
E.A. Kuznetsov, A.V. Mikhailov, Stability of stationary waves in nonlinear weakly dispersive
media. \textit{Sov. Phys. JETP} 40 (1975) 855-859.
\bibitem{HMP2023}
M.A. Hoefer, A. Mucalica, D.E. Pelinovsky, KdV breather on a cnoidal wave background,
\textit{J. Phys. A: Math. Theor.} 56 (2023) 185701.
\bibitem{Mau1999}
G.A. Maugin, \textit{Nonlinear Waves in Elastic Crystals}, Oxford University Press, Oxford, 1999.
\bibitem{S2001}
A.M. Samsonov, \textit{Strain Solitons in Solids and How to Construct Them}, Chapman \& Hall/CRC, Boca Raton, 2001.
\bibitem{P2003}
A.V. Porubov, \textit{Amplification of Nonlinear Strain Waves in Solids}, World Scientific, Singapore, 2003.
\bibitem{KS2008}
K.R. Khusnutdinova, A.M. Samsonov, Fission of a longitudinal strain solitary wave in a delaminated bar, \textit{Phys. Rev. E} 77 (2008) 066603.
\bibitem{EST2011}
J. Engelbrecht, A. Salupere, K. Tamm, Waves in microstructured solids and the Boussinesq paradigm, \textit{Wave Motion} 48 (2011) 717-726.
\bibitem{PTE2017}
T. Peets, K. Tamm, J. Engelbrecht, On the role of nonlinearities in the Boussinesq-type wave equations, \textit{Wave Motion} 71 (2017) 113-119.
\bibitem{W1974}
G. B. Whitham, \textit{Linear and Nonlinear Waves},  John Wiley \& Sons, New York, 1974.
\bibitem{J1997}
R. S. Johnson, \textit{A Modern Introduction to the Mathematical Theory of Water Waves}, Cambridge University Press, Cambridge, 1997.
\bibitem{HWTCH2022}
Y. He, A. Witt, S. Trillo, A. Chabchoub, N. Hoffmann,  Extreme wave excitation from localized phase-shift perturbations. \textit{Phys. Rev. E} 106  (2022) L043101.
\bibitem{SO1998}
T.P. Stanton, L.A. Ostrovsky, Observations of highly nonlinear internal solitons over the continental shelf, {\it Geoph. Res. Lett.}  25 (1998) 2695-2698.
\bibitem{CF1954}
C. Fox, An \textit{Introduction to the Calculus of Variations}, Oxford University Press, Oxford, 1954.



\end{thebibliography}
\end{document}